\documentclass{cplslarge}%
\usepackage[authoryear]{natbib}

\usepackage[switch]{lineno}

\usepackage[bookmarks = true, bookmarksnumbered = true, pdfpagemode =None, pdfstartview = FitH, pdfpagelayout = SinglePage, colorlinks = true, urlcolor = blue, citecolor = blue]{hyperref}

\usepackage{graphicx}

\usepackage{amsmath}
\usepackage{amssymb}
\usepackage{gensymb}
\usepackage{rotating}
\usepackage{lscape}

\global\chapterreferencefalse
\usepackage[sectionbib]{chapterbib}

\DeclareGraphicsExtensions{.eps,.jpg}

\begin{document}
\frontmatter
\maketitle

\mainmatter
  
\author[Sayanagi et al.]{k.m. sayanagi$^1$, k.h. baines$^2$, u.a. dyudina$^3$, l.n. fletcher$^4$, a. S{\' a}nchez-Lavega$^5$, r.a. west$^2$}

\chapter{Saturn's Polar Atmosphere}


\noindent $^1$Department of Atmospheric and Planetary Sciences, 
Hampton University, 23 E Tyler St., Hampton, VA 23668, USA

\noindent $^2$Jet Propulsion Laboratory, California Institute of Technology, 4800 Oak Grove Drive, Pasadena, CA 91109, USA

\noindent $^3$Division of Geological and Planetary Sciences, California Institute of Technology, 1200 California Blvd, Pasadena, CA 91125, USA

\noindent $^4$Department of Physics and Astronomy, University of Leicester, University Road, Leicester, LE1 7RH

\noindent $^5$Departamento de F\'{\i}sica Aplicada I, E.T.S. Ingenier\'{\i}a, Universidad del Pa\'{\i}s Vasco, Bilbao, Spain






\bigskip

\noindent \textbf{Copyright Notice}

\noindent The Chapter, Saturn's Polar Atmosphere, is to be published by Cambridge University Press as part of a multi-volume work edited by Kevin Baines, Michael Flasar, Norbert Krupp, and Thomas Stallard, entitled ``Saturn in the 21st Century'' (`the Volume').

\noindent \copyright~in the Chapter, K.M. Sayanagi, K.H. Baines, U.A. Dyudina, L.N. Fletcher, A. S{\' a}nchez-Lavega, and R.A. West

\noindent \copyright~in the Volume, Cambridge University Press 

\noindent NB: The copy of the Chapter, as displayed on this website, is a draft, pre-publication copy only. The final, published version of the Chapter will be available to purchase through Cambridge University Press and other standard distribution channels as part of the wider, edited Volume, once published. This draft copy is made available for personal use only and must not be sold or re-distributed.


\bigskip

\noindent \textbf{Abstract}

\noindent This chapter reviews the state of our knowledge about Saturn's polar atmosphere that has been revealed through Earth- and space-based observation as well as theoretical and numerical modeling. In particular, the Cassini mission to Saturn, which has been in orbit around the ringed planet since 2004, has revolutionized our understanding of the planet. The current review updates a previous review by \cite{DelGenio_etal_2009SatBook}, written after Cassini's primary mission phase that ended in 2008, by focusing on the north polar region of Saturn and comparing it to the southern high latitudes. Two prominent features in the northern high latitudes are the northern hexagon and the north polar vortex; we extensively review observational and theoretical investigations to date of both features. We also review the seasonal evolution of the polar regions using the observational data accumulated during the Cassini mission since 2004 (shortly after the northern winter solstice in 2002), through the equinox in 2009, and approaching the next solstice in 2017. We conclude the current review by listing unanswered questions and describing the observations of the polar regions planned for the \emph{Grand Finale} phase of the Cassini mission between 2016 and 2017. 

\section{Introduction: views of Saturn's poles from the Cassini orbiter}
\label{section:intro}
The Cassini mission has explored the Saturnian system in ways that are possible only from an orbiting spacecraft. Cassini is the fourth spacecraft to visit Saturn after Pioneer~11 and Voyagers 1 \& 2, and the first spacecraft to orbit the planet. Cassini has been in orbit around Saturn since 2004 and accumulated a decade of near-continuous observational data of the planet as of this writing, and will capture almost half a Saturn-year by the planned end of the mission in 2017. 

Among the new data returned from Cassini, its observations of Saturn's polar regions in particular have opened a new era in atmospheric studies of the giant planets. The data returned from the Cassini mission represent a significant advance in two major ways. The first is the long temporal coverage spanning nearly two solstices that has captured diverse atmospheric phenomena including seasonal changes. In comparison, each of the previous flyby missions captured data for only several months, which essentially represent temporal snapshots in the context of planetary climatology on a planet for which the orbital period is 30~Earth years. Second, Cassini's dynamic orbital trajectory has allowed the spacecraft to tour the Saturnian space between the equatorial plane to high above the poles, enabling observations of every latitude of Saturn from a wide range of illumination and observation angles. From those various observational geometries, infrared observations coupled with stellar occultation and radio occultation measurements can also provide a crucial vertical dimension to our understanding of the polar regions, extending hundreds of kilometers above the visible cloud deck and tropospheric hazes. These advantages have enabled detailed studies of both poles of Saturn, revealing surprising new results in the nature of polar winds, vortex features, aerosol properties and thermal structure, as we detail in this review. 

\subsection{Context of the current review}
This review of Saturn's poles integrates the most recent Cassini data, presenting multi-faceted results on cloud morphology, thermal structure, aerosol properties, and atmospheric dynamics. As such, this review builds on and extends previous work based on the Voyager flyby missions (e.g., \citealt{Gehrels_Matthews_1984SatBook}; in particular, \citealt{Tomasko_etal_1984} and \citealt{Ingersoll_etal_1984}), and on the first 5~years of the Cassini mission (\citealt{Dougherty_etal_2009SatBook}; specifically the reviews by \citealt{DelGenio_etal_2009SatBook} and \citealt{West_etal_2009SatBook}). Reviews of Jupiter's atmosphere also provide important points of comparison for the current review. The book by \cite{Gehrels_1976JupBook} contains reviews of Jupiter's polar atmosphere using the Pioneer~11 Infrared Radiometer data \citep{Ingersoll_etal_1976JupBook, Orton_Ingersoll_1976JupBook}. Another review book on Jupiter by \cite{Bagenal_etal_2004JupBook} treats the state of the knowledge after the Galileo mission \citep{West_etal_2004, Ingersoll_etal_2004book, Moses_etal_2004JupBook}. In addition, \cite{Dowling_1995AnnRev} and \cite{Vasavada_Showman_2005} provide in-depth reviews of processes related to jetstream formation and maintenance in giant planet atmospheres, and form the basis of our review of giant planet polar atmospheric dynamics. A textbook by \cite{Sanchez-Lavega_2011book} also provides a thorough introductory treatment of physical and chemical processes relevant to planetary atmospheres.

The Cassini observations of Saturn's poles represent our most in-depth polar dataset for a giant planet. The picture revealed for Saturn's poles provides a key point of comparison for Jupiter, Uranus and Neptune, at a critical time when knowledge of the polar environments of these planets is rapidly growing. 

From Earth, Jupiter's poles are not visible because the planet orbits in the ecliptic plane and its rotation axis has virtually no tilt; consequently, the poles can be viewed adequately only by spacecraft flying over the non-equatorial region of the planet. To date, only Pioneer 11 and Ulysses have departed the ecliptic significantly, and only the former carried an imaging instrument. In 2016, NASA's Juno spacecraft will enter orbits that pass directly over Jupiter's north and south poles, exploring these environments in detail for the first time. 

Uranus, with its rotation axis tilted by 98\degree, offers a unique opportunity to observe its poles from Earth. In the images of Uranus returned by Voyager~2, only a small number of discrete features were seen at latitudes poleward of 45\degree S latitude \citep{Allison_etal_1991UraBook, West_etal_1991UraBook}; however, modern digital image processing analysis by \cite{Karkoschka_2015_Uranus-V2} revealed many more features of Uranus near the south pole for the first time. HST and ground based observations leading up to Uranus' equinox were similarly free of discrete cloud features in the southern high latitudes (e.g., \citealt{Sromovsky_etal_2012_UraSpots}). In contrast, in the northern hemisphere, a large number of discrete features have been observed north of 45\degree N in ground based images \citep{Sromovsky_etal_2009_UraClouds, Sromovsky_etal_2012_UraSpots, Sromovsky_etal_2012_UraPostEquinox}. Continued studies of Uranus' hemispheric asymmetry, polar winds, and the north polar clouds are providing new insights into this planet's polar circulation (e.g. \citealt{Sromovsky_etal_2015_UranusCirc}). 

Neptune's polar environments may be most analogous to Saturn's in terms of seasonal effects because their obliquities are similar; Neptune's obliquity is 28.32\degree. Indeed, several observations of Neptune's south pole have resulted in comparisons with Saturn (e.g., \citealt{Hammel_etal_2007_NepEthaneMethane, Orton_etal_2007_NepSeasons}). In particular, Neptune's hotspot near the south pole, revealed by large ground-based telescopes armed with adaptive optics (e.g., \citealt{Luszcz-Cook_etal_2010Nep, Orton_etal_2012NepPole, Fletcher_etal_2014Nep, dePater_etal_2014NepCirc}), appears to be similar to those on Saturn (e.g., \citealt{Fletcher_etal_2008_CIRS_PoleHex}, and this chapter). 

As our knowledge of the poles of Jupiter, Uranus and Neptune continues to grow, the rich Saturn dataset returned by Cassini, provides the framework for understanding the polar environments of all four giant planets.

\subsection{Summary of Cassini polar datasets}
In the current review, the primary sources of data are the remote-sensing instruments onboard the Cassini spacecraft that cover the wavelengths between the ultraviolet and the thermal infrared. Here, we provide brief descriptions of the instruments that have played key roles in the studies presented in this review.

\begin{figure*}%
\begin{center}
\figurebox{6.85in}{}{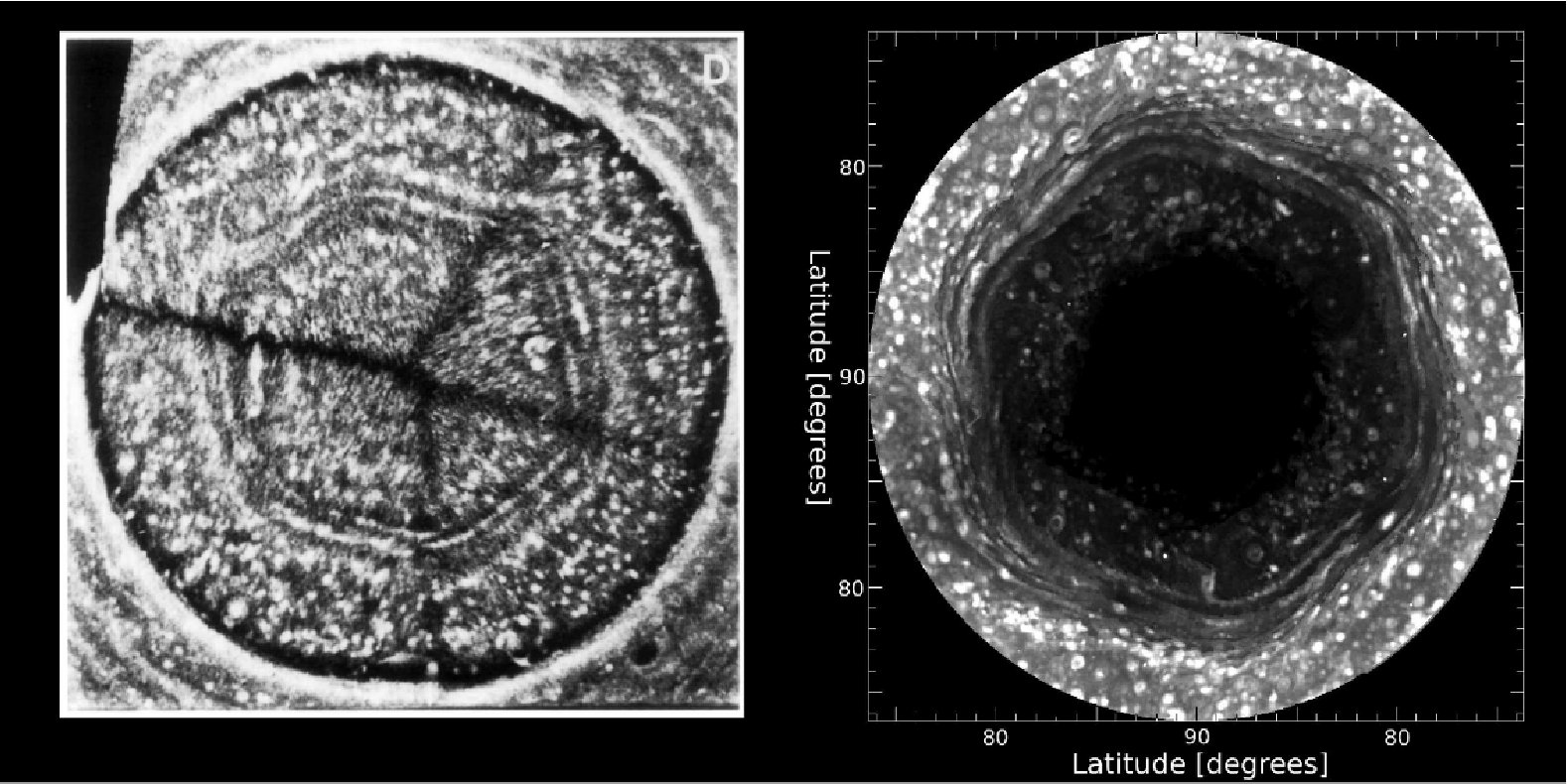}
\caption{Polar-projected mosaic of Saturn's north pole. The left map was generated by \cite{Godfrey_1988} using Voyager~2 images captured in August~1981 using the green filter. The right map was generated using images captured by Cassini ISS camera using the CB2 filter on January~3, 2009. Originally released by NASA as PIA11682.}
\label{Fig:Hex1609}
\end{center}
\end{figure*}

There are two infrared instruments that have played key roles. First, the Composite Infrared Spectrometer (CIRS; \citealt{Flasar_etal_2004_CIRS-SSRv}) is a Fourier transform spectrometer that provides spatially resolved spectral data at thermal infrared wavelengths between 7~$\mu$m and 1~mm; for the purpose of observing Saturn's atmosphere, the data returned by the instrument allow determinations of atmospheric thermal structure as well as chemical composition. The other infrared instrument, the Visible and Infrared Mapping Spectrometer (VIMS; \citealt{Brown_etal_2004_VIMS-SSRv}), is also an imaging spectrometer and covers the wavelength range between 0.3 and 5.1~$\mu$m. In particular, images captured in its 5-$\mu$m channel reveal clouds silhouetted against the background thermal emissions; the cloud features that appear in the VIMS 5-$\mu$m channel likely reside at the depth of several bars \citep{Baines_etal_2005b}. The 5-$\mu$m images' spatial resolutions are often high enough to track some cloud features to determine the wind fields (e.g., \citealt{Choi_Showman_Brown_2009} and \citealt{Baines_etal_2009poles}). 
High-resolution images in wavelengths between near-infrared and near-ultraviolet are captured using the Imaging Science Subsystem (ISS; \citealt{Porco_etal_2004}). ISS consists of two framing cameras with Charge-Coupled Device (CCD) detectors; the Narrow-Angle Camera (NAC) and the Wide-Angle Camera (WAC). The field of view (FOV) of WAC is ten times wider than that of NAC. Both cameras have multiple spectral filters; several of them have been used extensively in polar atmospheric measurements. In this review, we refer to these filters with their filter IDs as follows. The continuum band filters with central wavelengths of 750~nm (filter ID: CB2) and 935~nm (CB3) sense in the spectral windows between methane absorption bands, and show the top of the deep tropospheric cloud deck. The methane absorption band filters with central wavelengths of 727~nm (MT2) and 889~nm (MT3) are used to image the upper tropospheric haze layer that is believed to reside around the 100-150~mbar level. The broadband color filters with central wavelengths of 420~nm (VIO), 460~nm (BL1), 567~nm (GRN), and 648~nm (RED) are used to capture the colors of the clouds and hazes, which are indicative of compositional information. Each of these filters can be combined with a polarizing filter to analyze the properties of cloud and haze particles. At ultraviolet wavelengths, the Ultraviolet Imaging Spectrograph (UVIS; \citealt{Esposito_etal_2004}) captures spectrally resolved images in wavelengths between 56 and 190~nm that can be used to study the stratospheric aerosol properties and photochemical processes.


ISS and UVIS were able to image the south pole until the arrival of South Polar night beginning with the 2009 equinox. As Saturn approached the equinox, the northern hemisphere gradually emerged from the winter polar night and the extensive shadows cast by the rings. ISS and UVIS did not start returning images of the north pole until late 2012. As CIRS and VIMS are sensitive to thermal emissions of Saturn, they can resolve features in the winter polar night when the region is not illuminated by sunlight. The north pole has been illuminated since the 2009 equinox; however, the Cassini spacecraft was then in the equatorial plane where it had a poor view of the poles. The first post-equinoctial data of the poles were returned in late 2012 when the spacecraft raised its orbital inclination.

\subsection{Scope and organization of the current review}
This chapter first provides an overview of recent observations of the north polar region of Saturn, devoting sections to two of the major features that have received extensive observational coverage, namely, the northern hexagon (in Section~\ref{section:hex}) and the north polar vortex (in Section~\ref{section:npv}). We compare the north polar region to its southern counterpart in Section~\ref{section:n-s-compare}, updating the previous review by \cite{DelGenio_etal_2009SatBook}. We then focus on the physical processes that operate in the polar regions. Section~\ref{section:strat} discusses the thermal structure, seasonal forcing, and chemical processes in the polar stratosphere. The properties of the polar clouds and hazes are described in Section~\ref{section:aerosols}. A review of polar atmospheric dynamics theories and modeling are presented in Section~\ref{section:dynamics}. Finally, Section~\ref{section:conclusions} concludes the chapter with a summary of unanswered questions, and an outlook of the data expected to be returned by Cassini during the final years of the mission.

\adjustfigure{50pt}

\section{Northern hexagon}
\label{section:hex}

\subsection{Observations of the hexagon}
The hexagonal pattern in the cloud morphology that encircles the north pole of Saturn is a distinct feature that is not found on any of the other solar system planets; in this chapter, we refer to the feature simply as the hexagon. The feature was discovered by \cite{Godfrey_1988} in images taken at visible wavelengths during the Voyager~2 flyby of Saturn in 1981 (Fig.~\ref{Fig:Hex1609} left panel). The hexagon went unnoticed in these images for years after they were captured because the Voyager spacecraft had poor views of the poles due to their near-equatorial trajectories; the hexagon was discovered when the images were mosaicked and mapped in polar-projections by \cite{Godfrey_1988}.

\begin{figure}%
 \begin{center}
 \figurebox{3.4in}{}{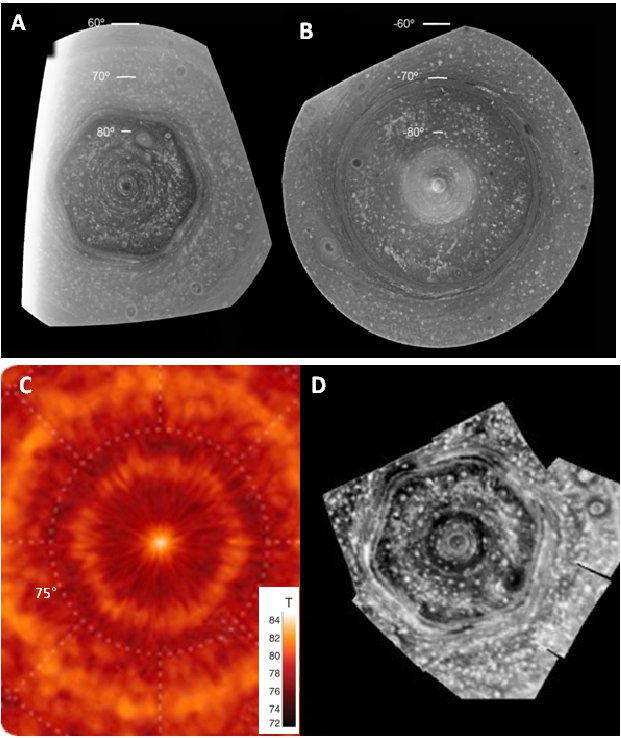}
 \caption{Polar-projected maps of Saturn's poles seen in three different Cassini imaging instruments. Panels A and B present the morphology of north and south polar regions, respectively, seen by the ISS camera using the CB2 (750~nm) filter \citep{Antunano_etal_2015_SatPoleDyn}. Panel C shows the temperature at the 100-mbar level retrieved from the CIRS spectra near 600 cm$^{-1}$ that is sensitive to the atmospheric temperatures near the tropopause \citep{Fletcher_etal_2008_CIRS_PoleHex}. Panel D shows the VIMS 5-$\mu$m thermal radiation map \citep{Baines_etal_2009poles}; the contrast is inverted in this panel such that high radiance is black and low radiance is bright, rendering clouds white like that in the ISS images of Panels A and B.}
 \label{Fig:HexMorph}
 \end{center}
 \end{figure}
\begin{figure}%
 \begin{center}
 \figurebox{3.2in}{}{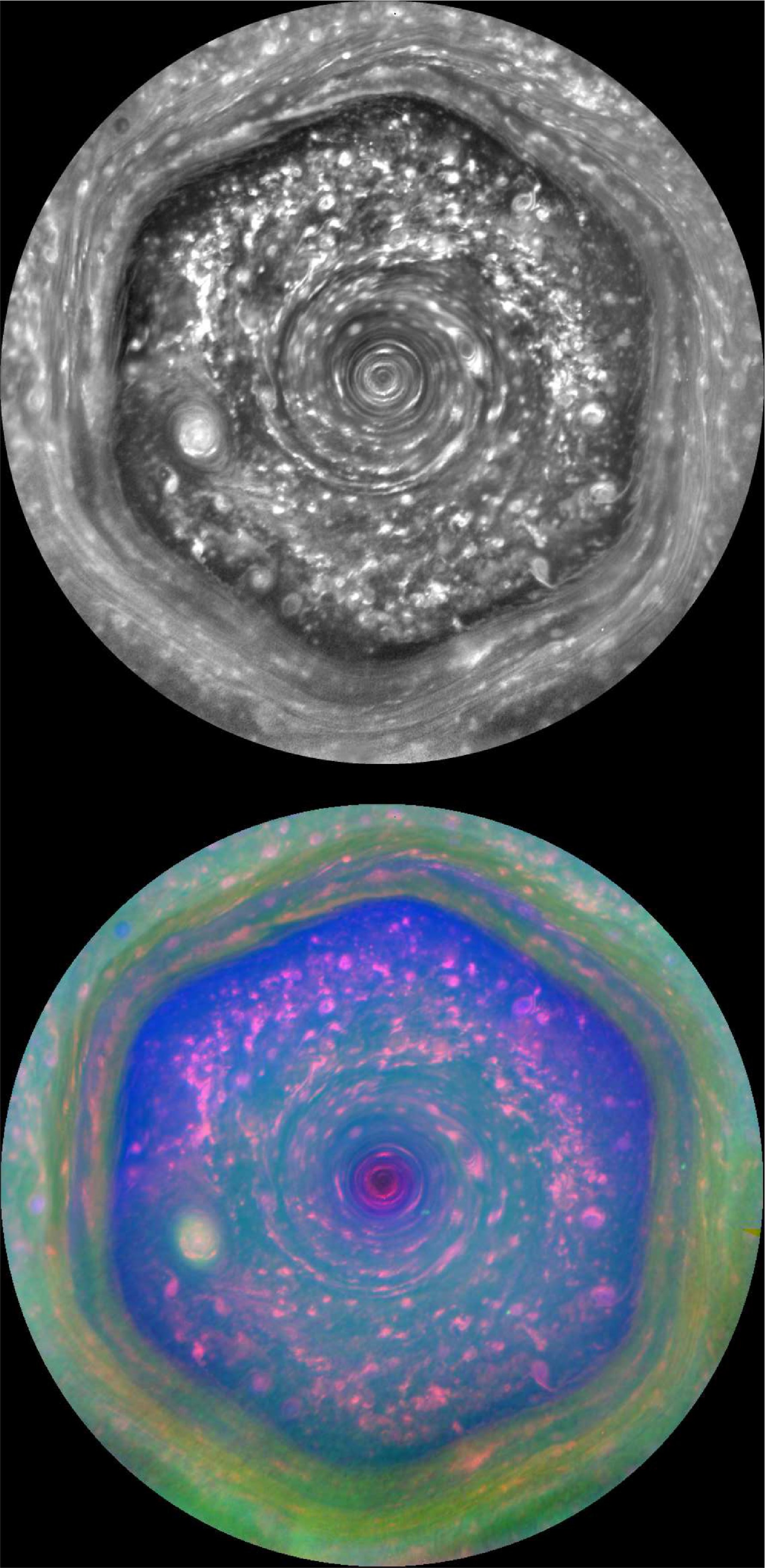}
 \caption{Polar-projected mosaic of Saturn's north pole captured by Cassini ISS camera on December 10, 2012. The images were obtained at a distance of approximately 533,304~km from Saturn. The smallest resolved features at the latitude of the hexagon have a horizontal scale of approximately 30 kilometer. The top panel is CB2 view. The bottom panel is a color composite which red channel is assigned to CB2, green is assigned to MT2 and blue channel is the sum of BL1 and VIO. Originally released by NASA as PIA17652.}
 \label{Fig:Hex1733}
 \end{center}
 \end{figure}

After the Voyagers, the hexagon was not detected again until after Saturn's north pole came into a favorable view from Earth around the solstice in 1987. The existence of the hexagon was confirmed in the images captured by the Hubble Space Telescope (HST) in 1991, which had been launched in 1990 \citep{Caldwell_etal_1993_HSTHEX}. The feature was also imaged by \cite{Sanchez-Lavega_etal_1993b} using ground based telescopes armed with the CCD imaging technology. \cite{Sanchez-Lavega_etal_1997hexagon} tracked the feature until the 1995 solstice when the pole began tilting away from the Earth, obscuring it from view. 
After the early 1990s, the north polar region was not observed again until the Cassini era. The orbiter arrived at Saturn in 2004; however, it did not have a good view of the poles until Cassini's first high-inclination mission phase between late 2006 and early 2007. During this period, the northern high latitudes were still shrouded in polar winter night, so it was the VIMS instrument that first confirmed the hexagon's presence using its 5-$\mu$m channel on October~29, 2006 (publicly released as PIA09188). The hexagonal shape was also recorded in the temperature map constructed using CIRS data captured on March~2, 2007 (Fig.~\ref{Fig:HexMorph}C; \citealt{Fletcher_etal_2008_CIRS_PoleHex}). In late 2008, sunlight started shining on the hexagon at 75\degree N planetocentric (PC) latitude, or 78\degree N planetographic (PG) latitude, which allowed visible-wavelength imaging using the ISS cameras; the first full-view of the hexagon was revealed in a polar mosaic captured on January~3, 2009 (Fig.~\ref{Fig:Hex1609} right panel) during another set of high-inclination orbits. The north pole has been illuminated by the sun since the equinox in August 2009; however, the next view of the hexagon had to wait until late 2012 because Cassini stayed in the equatorial plane during the intervening period. The first post-equinox set of images that showed the entire hexagon region was captured on December 10, 2012 (Fig.~\ref{Fig:Hex1733}). 

In summary, the hexagon was discovered in the Voyager images captured in 1980-1981, and confirmed using the HST and ground-based telescopes in 1991-1995. After the Cassini-Huygens spacecraft arrived at Saturn, it was seen in 2006-2007 using infrared instruments while the feature was still in winter polar darkness. After sunlight began to illuminate the feature in early 2008, the hexagon has been observed every year by the ISS cameras. Since 2012, the hexagon has also been clearly discernible in images captured by amateur astronomers armed with modest-sized telescopes (\citealt{Sanchez-Lavega_etal_2014_Hexagon}; \emph{c.f.} Chapter~14). Saturn's polar latitudes experience a strong seasonal forcing \citep{Perez-Hoyos_Sanchez-Lavega_2006}; even though there are large gaps in the observational records of the northern high-latitudes of Saturn, these observations clearly demonstrate that the hexagon is a long-lived structure that persists for longer than a Saturnian year.

 \begin{figure}%
 \begin{center}
 \figurebox{3.4in}{}{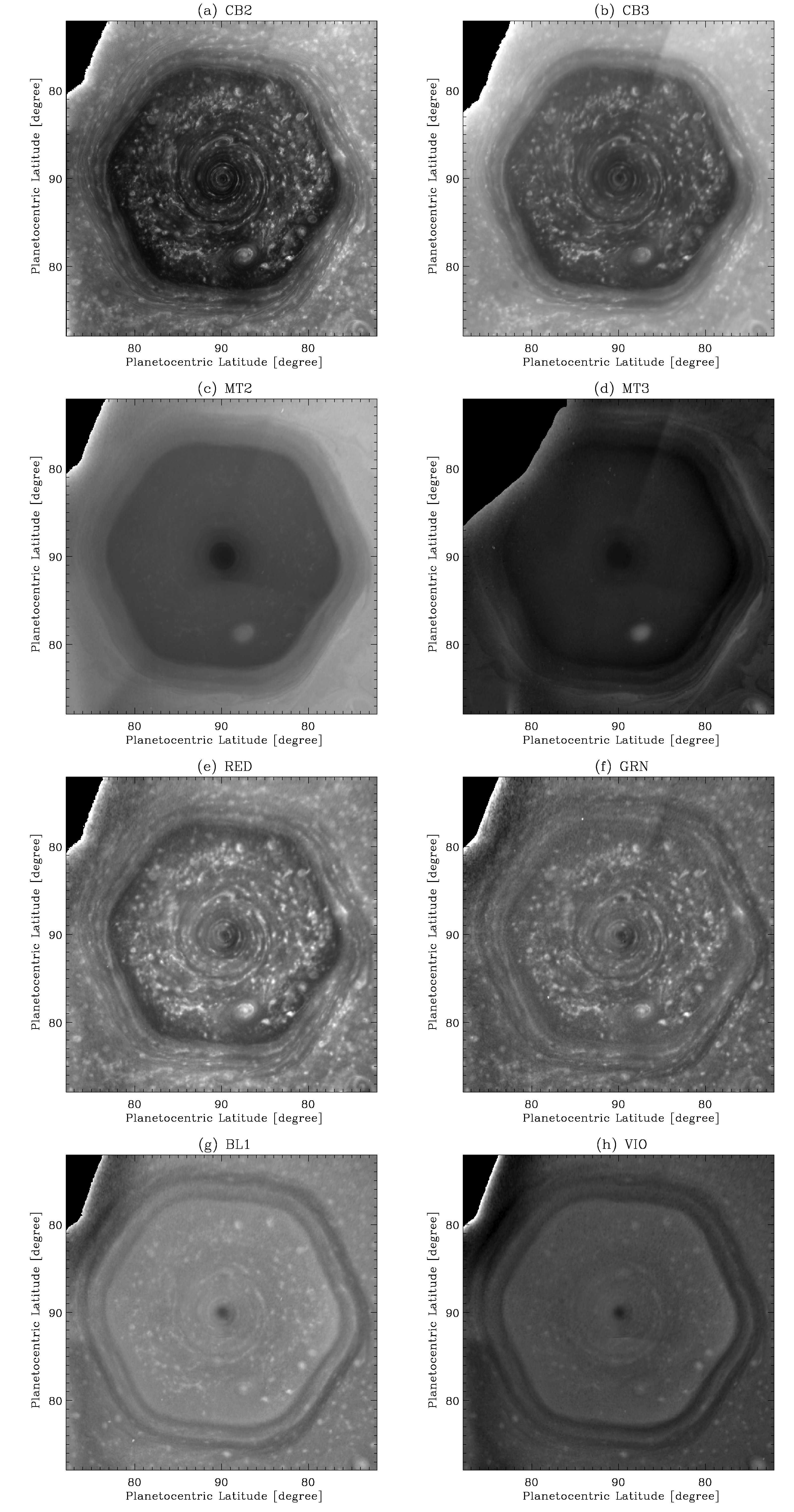}
 \caption{Cassini ISS WAC images of northern high-latitudes in polar orthographic projection \citep{Sayanagi_etal_2015DPS}. The filter used to make each of the map is denoted in the panel header (see main text for their central wavelengths).}
\label{Fig:NPV_WAC}
 \end{center}
 \end{figure}

The hexagon is a unique feature that has been found only around the north pole of Saturn, although some of its morphological characteristics are shared by features on elsewhere in our solar system. For example, terrestrial hurricanes sometimes develop a polygonal pattern inside the eye \citep{Lewis_Hawkins_1982, Schubert_etal_1999HurricanePolygonalEyes, Kossin_Schubert_2001HurricanePolygon}. The polar vortex of Venus exhibits transient polygonal characteristics \citep{Limaye_etal_2009VenusPolarVort, Piccioni_etal_2007VenSpole}. Earth's mid-latitude jetstream in the winter hemisphere often develops a stationary meandering pattern (e.g., \citealt{Held_etal_2002StationaryWave}). On Saturn, the eastward jet at 55.2\degree N PC (60.5\degree S PG) has been seen to develop a transient wavy appearance \citep{Vasavada_etal_2006}. On Jupiter, almost every jet harbors waves; for example, the southern hemisphere jets at 49.3\degree S, 58.4\degree S, and 64.1\degree S PC latitudes (53.1\degree S, 61.7\degree S and 67.0\degree S PG latitudes, respectively) sometimes exhibit wavy cloud morphologies \citep{Sanchez-Lavega_Hueso_Acarreta_1998}. In the northern hemisphere of Jupiter, the westward jet at 27\degree N PC (30\degree N PG; \citealt{Cosentino_etal_2015JupRibbon}) and the eastward jet at 62\degree N PC (65\degree N PG; \citealt{Orton_etal_2003, Barrado-Izagirre_etal_2008}) latitudes also develop transient wavy morphologies. Although these features exhibit morphological similarities to Saturn's northern hexagon, the mechanisms that generate these morphological characteristics may not be the same.

\subsection{Cloud morphology}
The noticeable characteristic that distinguishes the hexagon is the six-sided shape of the feature that encircles the north pole at around 75\degree N PC (78\degree N PG) latitude. The hexagonal shape is approximately 30,000~km across centered on the north pole. This striking geometric shape is a pattern in the atmospheric reflectivity. 

In a latitude-longitude cylindrical projection, the hexagon's outline is very close to a simple sinusoidal function with zonal wavenumber-6 that meanders between 75.3\degree N and 76.3\degree N PC (78.0\degree N and 78.9\degree N PG) latitude. This sinusoidal amplitude is such that the vertices are formed by the latitudinal minima in the sinusoid (i.e., the point of maximum northward curvature) and the flat sides correspond to the latitudinal maxima (the point of maximum southward curvature) \citep{Antunano_etal_2015_SatPoleDyn}. \cite{Morales-Juberias_etal_2015Hexagon}'s numerical models showed that the sinusoidal amplitude can be tuned by the vertical wind shear as discussed further in Section~12.7.3. Note that the hexagonal morphology becomes noticeable only when the planet is viewed from a viewpoint above the north pole multiple Saturn radii away; an in-situ traditional meteorological measurement would first reveal a remarkably monochromatic wavenumber-6 atmospheric structure, but a hexagonal geometry would not be apparent until it is mapped in a polar orthographic projection.

The hexagonal shape stands out especially in the near-infrared continuum filters CB2 (Figs.~\ref{Fig:NPV_WAC}a and~\ref{Fig:HexMorph}A) and CB3 (Fig.~\ref{Fig:NPV_WAC}b), and the methane absorption filters MT2 and MT3 (Figs.~\ref{Fig:NPV_WAC}c and ~\ref{Fig:NPV_WAC}d). In those filters, the interior (i.e., the region poleward of the hexagon) appeared significantly darker than the exterior with a sharp, well-defined boundary. Long-term observation of the hexagon over 5~Earth-years show that the vertices of this boundary maintain a mean latitude of 74.7\degree N PC (77.4\degree N PG) \citep{Sanchez-Lavega_etal_2014_Hexagon}.

In the CB2 and CB3 view, the interior of the hexagon is filled with vortices of various sizes from the limit of image resolution to several thousand kilometers. In MT2 and MT3, most of those vortices are not visible except for the large anticyclone, appearing bright around the 5-o'clock position of Figs.~\ref{Fig:NPV_WAC}c and \ref{Fig:NPV_WAC}d; this large anticyclone has been seen since the arrival of Cassini at Saturn, but it was not seen in Voyager images captured in 1980-81 (\citealt{Godfrey_1988} found a large anticyclone \emph{outside} of the hexagon). In MT2 and MT3, the albedo darkening toward the pole happens in three steps; the outer two steps form two concentric hexagons that surround the hexagonal jet. The inner and outer edges of the hexagon are separated by approximately 2.8\degree of latitude \citep{Sanchez-Lavega_etal_2014_Hexagon}. The last step toward the pole is associated with the north polar vortex, to be discussed further in Section~\ref{section:npv}. 

In the visible wavelengths, the brightness contrast gradually reverses with wavelength from red to violet such that the interior of the hexagon becomes brighter relative to the exterior. In the RED filter (Fig.~\ref{Fig:NPV_WAC}e), the appearance is similar to the continuum filter; the interior of the hexagon is darker than the exterior. In GRN, the brightness of the interior and exterior are approximately equal, and the hexagon presents a muted appearance (Fig.~\ref{Fig:NPV_WAC}f). The interior of the hexagon is brighter than the exterior in BL1 and VIO (Figs.~\ref{Fig:NPV_WAC}g and~\ref{Fig:NPV_WAC}f). At even shorter wavelengths in ultraviolet, the interior of the hexagon outline is darker than the surrounding as shown in Fig.~\ref{Fig:UVIShex}. In BL1 and VIO, the hexagon shape consists of two parallel tracks of dark bands; these tracks roughly follow the inner and outer edges apparent in the MT2 and MT3 images. These double tracks are also apparent in the VIMS 5-$\mu$m images (Fig.~\ref{Fig:HexMorph}D), which sense the opacity structure of the cloud layer predominantly residing at around 1.5-4~bar. 

A long-term tracking from 1980 to 2014 of the hexagon's vertices in the System~III longitude shows the hexagon to be nearly stationary, propagating westward with a zonal velocity of 0.036~m~s$^{-1}$, or a longitudinal propagation rate of $\langle \omega \rangle$ = $-$0.0129$\pm$0.0020\degree~day$^{-1}$ (negative number denotes a westward motion, and the brackets $\langle ~ \rangle$ denote averaging in time). In the inertial reference frame, this translates to an absolute angular velocity of $\langle \Omega \rangle$ = 810.7810\degree~day$^{-1}$ or a rotation period of $\langle \tau \rangle$ = 10h39min23.01$\pm$0.01s \citep{Sanchez-Lavega_etal_2014_Hexagon}. Note that the System~III rotation rate is defined as the angular rotation rate of $\Omega_{\textrm{III}}$ = 810.7939024\degree~day$^{-1}$ by \cite{Seidelmann_etal_2007} and \cite{Archinal_etal_2011_PlanetCartography} based on the primary periodicity in the Saturn Kilometric Radiation (SKR) during the Voyager flybys \citep{Desch_Kaiser_1981}; however, the SKR period has been found to be variable in subsequent measurements (e.g. \citealt{Galopeau_Lecacheux_2000}, \citealt{Fischer_etal_2014StormSKR}). As such, there is currently no evidence that Saturn's deep interior rotates at the System~III rate derived from the Voyager-era SKR period. As of this writing, System~III continues to be used as the standard reference frame; however, this convention may change in the future.

\begin{figure*}%
\begin{center}
\figurebox{4.0in}{}{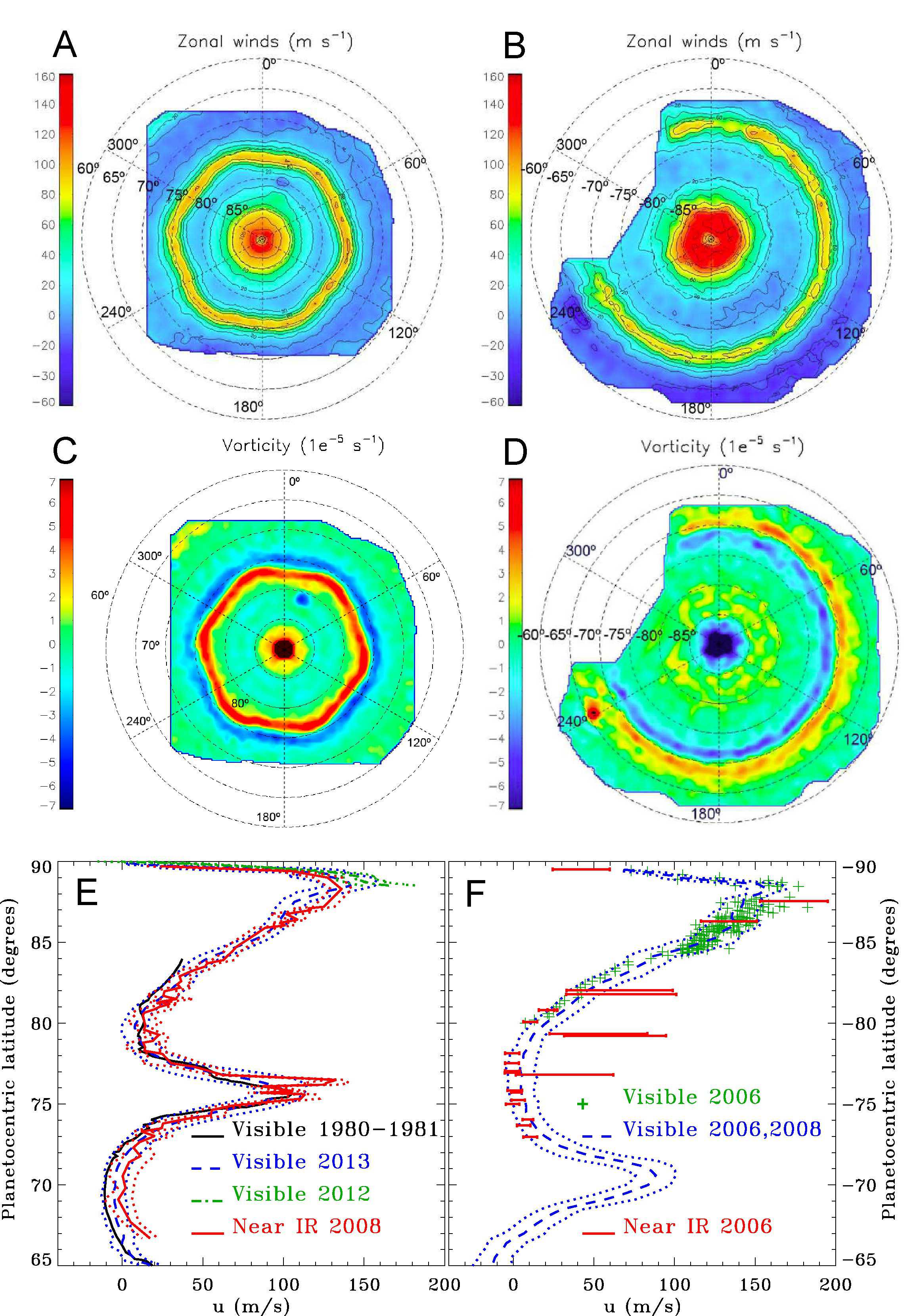}
\caption{\textbf{Panels A-D:} Maps of zonal wind speeds and vorticity. The panels in the left show measurements in the north polar region, and those in the right are for the south. Panels A and B show the zonal winds and panels C and D are vorticity measured from Cassini ISS images by \cite{Antunano_etal_2015_SatPoleDyn}. \textbf{Panels E-F:} Black solid line shows zonal winds measured by Voyager cameras during flybys in 1980-1981 \citep{Godfrey_1988}. Blue dashed lines with dotted lines denoting error bars show winds measured from Cassini ISS images in 2006, 2008 and 2013 \cite{Antunano_etal_2015_SatPoleDyn}. Green dot-dashed line with dotted lines denoting error bars in panel E shows measurements from Cassini ISS in 2012 \cite{Sayanagi_etal_2015npv}. Green + symbols in panel F are measurements from Cassini ISS in 2006 \cite{Dyudina_etal_2009}. Red lines are from Cassini VIMS 5-micron images in 2008 \cite{Baines_etal_2009poles} and 2006 \citep{Dyudina_etal_2009}.}
\label{Fig:HexZonalWind}
\end{center}
\end{figure*}

In the Voyager images, a large anticyclone that measured 11,000 km in the east-west dimension was seen lined up outside one of the hexagon edges \citep{Godfrey_1988} --- the vortex is identified as the North Polar Spot (NPS) in the literature. The NPS was re-observed by the HST \citep{Caldwell_etal_1993_HSTHEX} and ground-based images between 1990-95 \citep{Sanchez-Lavega_etal_1993b, Sanchez-Lavega_etal_1997hexagon}. In 1980-81 and 1990-95, the NPS drifted at a longitudinal rate of $\langle \omega \rangle$= 0.035\degree~day$^{-1}$ (i.e., eastward), and it was suggested to be dynamically coupled to the hexagon. \cite{Allison_Godfrey_Beebe_1990} proposed that the hexagon was a Rossby wave excited by the large vortex impinging on the jetstream; however, the NPS has not been present since Cassini arrived at Saturn in 2004, indicating that the hexagon can exist without the NPS. \cite{Sanchez-Lavega_etal_2014_Hexagon} interpret the hexagon as a stationary trapped three-dimensional unforced Rossby wave trapped in the jet, which acts as a waveguide. The trapped Rossby wave manifests as a meandering in the jet as demonstrated by \cite{Morales-Juberias_etal_2015Hexagon} as discussed further in Section~\ref{section:dynamics}.


Around the south pole of Saturn, there is no hexagon or other permanent polygons in the southern-most zonal jet at 70.4$\pm$0.1\degree S PC latitude (73.4\degree S PG) with a peak speed of $u$=87$\pm$12~m~s$^{-1}$, which can be considered the southern counterpart of the hexagon jet at 75\degree N PC (78\degree N PG), as documented using HST images \citep{Sanchez-Lavega_etal_2002_SatSouth} and Cassini ISS images \citep{Antunano_etal_2015_SatPoleDyn}. The southern jet does not follow a path that oscillates in latitude, and instead it follows the latitude circle (Fig.~\ref{Fig:HexZonalWind} panels E and F). Both jets have similar width and a maximum relative vorticity $\textrm{d}u/\textrm{d}y$ = 5$\times$10$^{-5}$~s$^{-1}$ which is about 10$^{-1}$ of the planetary vorticity (the Coriolis parameter at Saturn's $\sim$75\degree~latitude is $f\approx$3.2$\times$10$^{-4}$~s$^{-1}$). Although the southern-most jet does not harbor a polygonal wave,, a transient polygonal wave was noted in the eastward jet at 55.2\degree N PC (60.5\degree S PG) by \citealt{Vasavada_etal_2006}.

\subsection{Observed dynamical characteristics}
Voyager images revealed that small clouds were moving alongside the outline of the hexagon, showing that the hexagon is associated with a strong eastward jet \citep{Godfrey_1988}. Cassini ISS and VIMS images confirmed the Voyager observation \citep{Baines_etal_2009poles, Sanchez-Lavega_etal_2014_Hexagon, Antunano_etal_2015_SatPoleDyn}. The zonal wind structure is shown in Fig.~\ref{Fig:HexZonalWind}.

The jet is centered at 75.5\degree N PC (78.1\degree N PG) latitude and has a mean peak velocity $\bar{u}$ = 104$\pm$15~m~s$^{-1}$ with some features reaching 120~m~s$^{-1}$ and a meridional width (defined by the Gaussian FWHM) of $L_\textrm{jet}$ = 2,900 km (see Fig.~\ref{Fig:HexZonalWind}). The hexagon wind field shows a prominent wavenumber-6 structure in the meridional component of the wind with a magnitude of $v\sim$28~m~s$^{-1}$ --- this structure arises from the meridionally meandering path of the hexagon jet.

\adjustfigure{130pt}

\begin{table*}
\begin{center}
\caption{Hexagon atmospheric dynamics parameters determined by \cite{Sanchez-Lavega_etal_2014_Hexagon}.}
\begin{tabular} {l l l}
  \hline \hline
  Notation & Parameters & Value\\ 
  \hline
        $L_x$ & Zonal Length Scale & 14,500~km \\
        $L_y$ & Meridional Length Scale & 5,785~km \\
        $f_\mathrm{0}$ & Coriolis Parameter at Hexagon Jet Center & 3.17$\times$10$^{-4}$~s$^{-1}$ \\
        $\beta$ & Coriolis Gradient at Hexagon Jet Center & 1.5$\times$10$^{-12}$~m$^{-1}$s$^{-1}$ \\
        $c_x$ & Zonal Wave Phase Speed & $-$0.036~m~s$^{-1}$ \\
        $\bar{u}$ & Mean Hexagon Jet Speed & 120~m~s$^{-1}$ \\
        $\mathrm{d}^2 u /\mathrm{d}y^2$ & Zonal Wind Curvature & $-$6-8~$\times$10$^{-11}$~m$^{-1}$s$^{-1}$ \\
        $N_\mathrm{B}$ & Brunt-V\"{a}is\"{a}l\"{a} Frequency & 8.3$\times$10$^{-3}$ s$^{-1}$ \\
        $L_\mathrm{D}$ & Deformation Radius & $\sim$1000~km \\
        \hline \hline
 \end{tabular}
 \label{tab:hex_params}
\end{center}
\end{table*}

The hexagon is also detected in the thermal infrared in a wavelength sensitive to the upper tropospheric temperatures in the 100 - 800~mbar altitude; however, it is not apparent in the wavelength range of 7.4-8 $\mu$m sensitive to the stratosphere at the 1-6~mbar altitude level (\cite{Fletcher_etal_2008_CIRS_PoleHex}, Figs.~\ref{Fig:HexMorph}C and \ref{Fig:polarmaps} lower left panel). The temperature maps show the hexagon is manifested as a warm band at 77\degree N PC (79\degree N PG) with a cold band outside at 73\degree N PC (76\degree N PG), coincident with the 5~$\mu$m bright inner band (cloud clearing, cyclonic vorticity) and dark band (cloud covered, anticyclonic vorticity) on VIMS images respectively \citep{Baines_etal_2009poles}. This temperature - cloud opacity relationship has been explained as due to upwelling on the equator side and subsidence on the poleward side of the hexagon \citep{Fletcher_etal_2008_CIRS_PoleHex}. 

From the observed characteristics, \cite{Sanchez-Lavega_etal_2014_Hexagon} deduced the following dynamical parameters for the hexagon. The notation and the derivations follow those in Chapter~8 of \cite{Sanchez-Lavega_2011book}. The zonal, meridional and vertical wavenumbers $k$, $l$, and $m$, respectively, are defined as
\begin{equation} \label{e:klm_def}
k = \frac{2\pi}{L_x}\mathrm{;~}l = \frac{2\pi}{L_y}\textrm{;~}m = \frac{2\pi}{L_z}
\end{equation}
where $L_x$, $L_y$, and $L_z$ are characteristic zonal, meridional and vertical length scales, respectively. We take the zonal wavelength of the hexagon as the zonal length scale $L_x$, and the jet's width as the meridional length scale $L_y$. We take the measured value of the phase speed for $c_x$. The values of these dynamical parameters are summarized in Table~\ref{tab:hex_params}. Substituting the dynamical parameters into the three-dimensional Rossby wave dispersion relationship 
\begin{equation} \label{e:Rossby_Dispersion}
c_x -\bar{u} = -\frac{\beta -\textrm{d}^2 \bar{u}/\textrm{d}y^2}{k^2 +l^2 +\left( {f_0}^2/{N_\mathrm{B}}^2 \right)m^2 +1/4L_\mathrm{D}^2 },
\end{equation}
reveals that, for the wave to propagate, the vertical wavenumber must be imaginary, $m^2 < 0$. In the equation, $\bar{u}$ is the zonal mean wind profile, $\mathrm{d}^2 \bar{u} /\mathrm{d}y^2$ is the curvature of the zonal wind at the center of the jet, $y$ is the north-south distance, and $L_\mathrm{D} = N_\mathrm{B}H/f$ is the deformation radius with $H$ as the atmospheric scale height. $\beta$ is the gradient of the planetary vorticity; at the latitude of the hexagon, $\beta = \textrm{d}f/\textrm{d}y = 2 \Omega_\textrm{III} \cos{\varphi_{0}} / R_S =$1.5$\times$10$^{-12}$~m$^{-1}$s$^{-1}$ ($R_S$ = 55,250 km is the radius of local surface curvature on Saturn at the jet's latitude). The upper-limit for the Brunt-V\"{a}is\"{a}l\"{a} frequency $N_\mathrm{B}$ is estimated by \cite{Lindal_etal_1985} to be $N_\mathrm{B}\lesssim$ 10$^{-2}$~s$^{-1}$; \cite{Sanchez-Lavega_etal_2014_Hexagon} used $N_\mathrm{B}$=8.3$\times$10$^{-3}$ s$^{-1}$. This suggests that the Rossby wave is vertically trapped within the region of positive static stability in Saturn's atmosphere, which is in agreement with a previous analysis by \cite{Allison_Godfrey_Beebe_1990}. The depth of the layer that is trapping the Rossby wave is unknown --- the layer could extend well below the water condensation level at $\sim$10-bar; however, such interpretation would contradict \cite{Morales-Juberias_etal_2015Hexagon}'s results as discussed in Section~12.7.3. 

\begin{figure*}%
\begin{center}
\figurebox{6.5in}{}{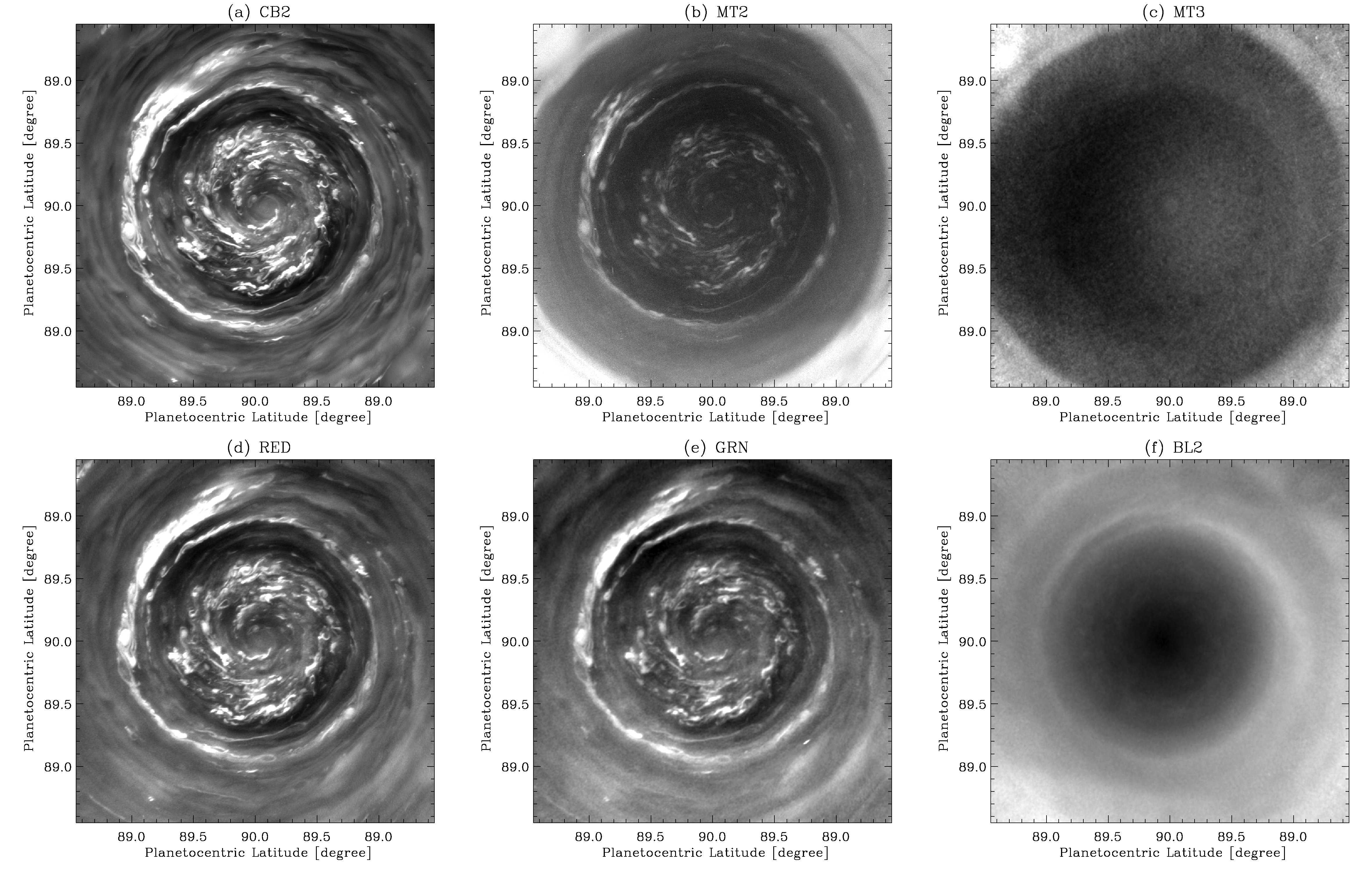}
\caption{Cassini ISS NAC images of the northern high-latitudes in polar orthographic projection. The filter used to make each of the map is denoted in the panel header. From \cite{Sayanagi_etal_2015npv}.}
\label{Fig:NPV_NAC}
\end{center}
\end{figure*}

The lower limit to the vertical wavenumber $m$ (or, upper limit to the vertical length scale $L_z$) can be estimated by deriving the condition for which the vertical wavenumber is real, $m^2 > 0$ (i.e., the wave is vertically propagating with a real wavelength $L_z$). The resulting condition demands a meridional length scale of $L_y >$15,700~km, which is not consistent with the observed characteristics of the hexagon; a jet centered at 75\degree N PC (78\degree N PG) with a meridional width of $L_y$=15,700~km would overtake the pole to the north and the neighboring jet to the south. As noted earlier, Saturn's internal rotation rate remains uncertain; however, note that above interpretation is not affected by the uncertainties in the rotation rate because $c_x -\bar{u}$ is invariant to the rotation rate except for a minor dependence on $f_0$ and $\beta$. Since $c_x$ and $\bar{u}$ are both observed in a common reference frame of System~III, shifting them to another common reference frame with a different rotation rate does not substantially affect $c_x -\bar{u}$. Furthermore, note that the potential vorticity is invariant to the planetary rotation; thus, we do not expect that a small uncertainty in the planet's rotation rate would significantly affect atmospheric dynamics.

The eastward hexagon jet centered at 75\degree N PC (78\degree N PG) is in geostrophic balance, indicated by the Rossby number $R_\mathrm{o}=u/f_0 L \sim$0.13. The jet violates the Rayleigh-Kuo stability criterion, which is a necessary but not sufficient condition that states that a flow is stable if the gradient of the absolute vorticity $\beta -\mathrm{d^2}\bar{u}/\mathrm{d}y^2$ does not change sign anywhere in the flow; the condition is applicable to a geostrophic barotropic zonal jet \citep{Kuo_1949}. The zonal wind curvature is negative at the peak of the jet, $\textrm{d}^{2} \bar{u} / \textrm{d}y^2 \sim -7\times10^{-11}$ ~m$^{-1}$s$^{-1}$ $< 0$, so the absolute vorticity gradient is positive $\beta -\mathrm{d^2}\bar{u}/\mathrm{d}y^2 > 0$. The zonal wind curvature has maxima to the south of the jet at 74\degree N PC (77\degree N PG) $\textrm{d}^{2} \bar{u} / \textrm{d}y^2 \sim 2\times10^{-11}$~m$^{-1}$s$^{-1}$ and to the north at 78\degree N PC (80\degree N PG) $\textrm{d}^{2} \bar{u} / \textrm{d}y^2 \sim 4\times10^{-11}$~m$^{-1}$s$^{-1}$, both of which can over come the local planetary vorticity gradient $\beta \sim 1.5 \times$10$^{-12}$~m$^{-1}$s$^{-1}$ \citep{Antunano_etal_2015_SatPoleDyn}. Thus $\beta -\mathrm{d}^{2}\bar{u}/\mathrm{d}y^{2}$ changes sign in the flanks of the hexagon jet and violates the Rayleigh-Kuo criterion. The jets on Saturn also violate the Charney-Stern criterion (a jet is stable when the gradient of potential vorticity does not change sign; \citealt{Charney_Stern_1962, Read_etal_2009}). Saturnian jets may be neutrally stable according to Arnol'd's Second Criterion \citep{Read_etal_2009, Read_Dowling_Schubert_2009},
\begin{equation} \label{e:A-II_criterion}
\frac{\bar{u} -c_x}{\partial \bar{q}/ \partial y} \geq L_\mathrm{D}^2,
\end{equation}
where $c_x$ is the fastest propagating wave's phase speed and $\bar{q}$ is the zonal mean potential vorticity. The criterion states that, for the wind to be stable, the magnitude of the wind speed must be greater than the phase speed of the fastest up-stream propagating wave by $L_\mathrm{D}^2 \partial \bar{q}/ \partial y$ \citep{Arnold_1966, Dowling_2014_Arnold-II}. \cite{Dowling_1995AnnRev} and \cite{Vallis_2006} discuss these stability criteria. The depth of the jet structure and the stability of the zonal jets on Saturn are related and still need to be explained (to be discussed further in Section~\ref{section:dynamics}). These violations of stability criteria are not unique to Saturn's atmospheric jets; the forcing and maintenance mechanisms of giant planet atmospheric zonal jets is a continuing topic of research as also discussed in Chapter~11.

\section{North polar vortex}
\label{section:npv}
\subsection{Cloud morphology}
The north polar vortex (NPV) is an intense cyclonic vortex revealed in the Cassini era that resides inside of the hexagon and centered on the north pole. The presence of the NPV was was detected in CIRS data by \cite{Fletcher_etal_2008_CIRS_PoleHex}. The cloud morphology of the NPV was first documented by \cite{Baines_etal_2009poles} using VIMS 5-$\mu$m images captured in 2008 (which can be seen at the center of Fig.~\ref{Fig:HexMorph}D) before the 2009 equinox while the north pole was still under the winter polar darkness. The VIMS observation by \cite{Baines_etal_2009poles} also detected cloud motions and provided a direct confirmation that there is an intense cyclonic vortex over the north pole of Saturn (Fig.~\ref{Fig:HexZonalWind}E). The cyclonic cloud motions have also been shown later by cloud tracking in Cassini ISS visible observations after the Sun illuminated the North Pole following the 2009 equinox \citep{Antunano_etal_2015_SatPoleDyn, Sayanagi_etal_2015npv}. The ISS observations also show that there is a large spiraling cloud structure centered on the north pole as can be seen in Fig.~\ref{Fig:NPV_WAC}. The sense of the spiral was cyclonic (i.e., a spiraling arm traces a line that turns clockwise as it moves away from the pole), which is consistent with the thermal hotspot revealed by \cite{Fletcher_etal_2008_CIRS_PoleHex}. The south-polar vortex also has cyclonic vorticity \citep{Sanchez-Lavega_etal_2006, Dyudina_etal_2008, Dyudina_etal_2009}.

\adjustfigure{20pt}

\begin{figure*}%
\begin{center}
\figurebox{6.85in}{}{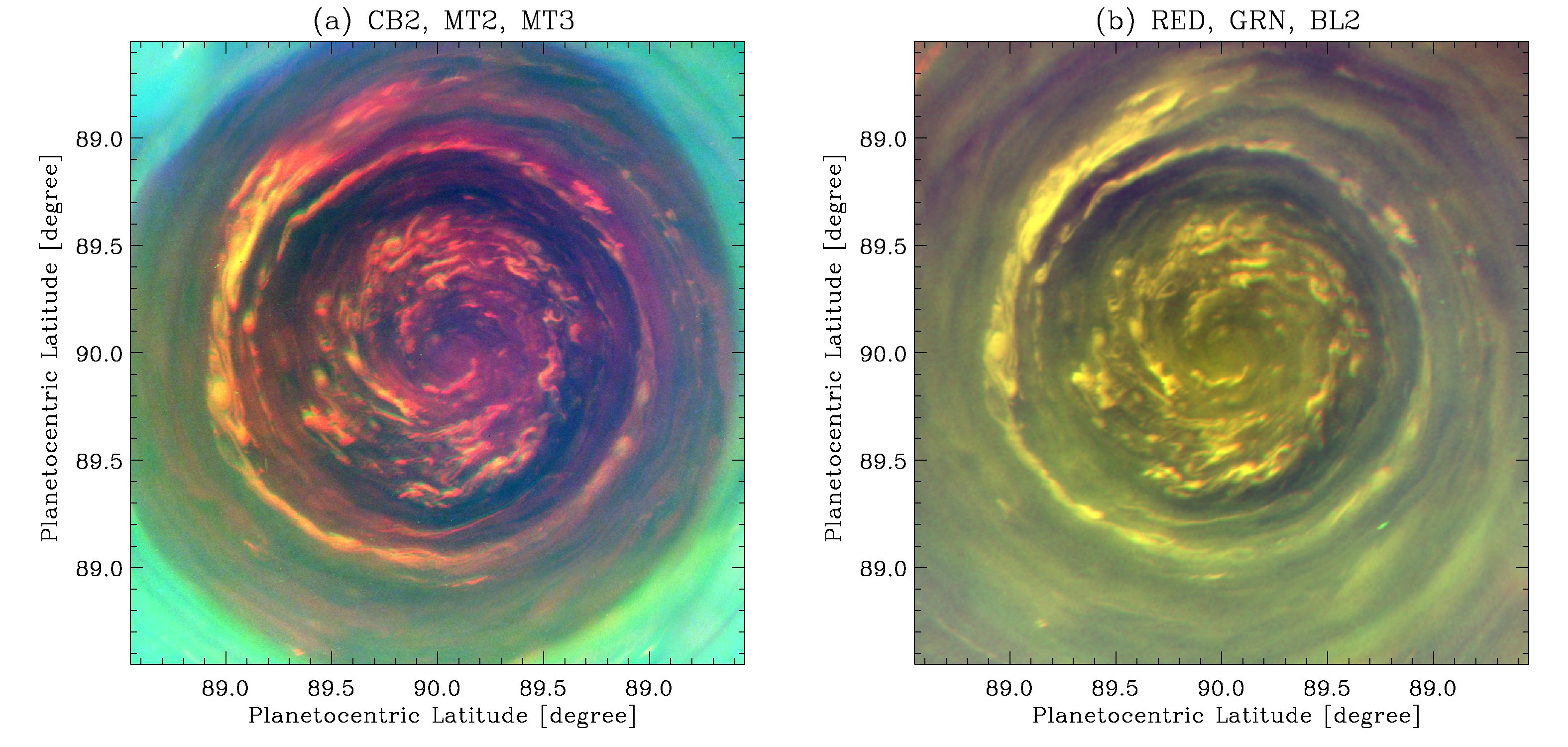}
\caption{Color composites of northern high-latitudes in polar orthographic projection. The red, green and blue channels of the composites are assigned to the filters identified in each of the panel headers. From \cite{Sayanagi_etal_2015npv}.}
\label{Fig:NPV_color}
\end{center}
\end{figure*}

The morphological context that surrounds the polar vortex is presented in Fig.~\ref{Fig:NPV_WAC}. The high-resolution cloud and wind structure of the north polar vortex are captured in ISS visible images \citep{Antunano_etal_2015_SatPoleDyn, Sayanagi_etal_2015npv}; which yield higher spatial resolution than the infrared VIMS images. The ISS instrument also returns images in multiple wavelength bands using different filters. Images captured in CB2 and CB3 show a large spiraling cloud morphology (Figs.~\ref{Fig:NPV_WAC}a and ~\ref{Fig:NPV_WAC}b). The NPV's spiraling appearance in RED (Fig.~\ref{Fig:NPV_WAC}e) and GRN (Fig.~\ref{Fig:NPV_WAC}f) are similar to that in the CB2. In BL1 and VIO (Figs.~\ref{Fig:NPV_WAC}g and ~\ref{Fig:NPV_WAC}h), the spiraling structure is barely discernible, and poleward of about 89.5\degree N PC (89.6\degree N PG) latitude is dark, appearing as a circular hole. In the methane absorption bands MT2 and MT3 (Figs.~\ref{Fig:NPV_WAC}c and~\ref{Fig:NPV_WAC}d), the contrast is dominated by the dark circular hole centered on the pole; the rim of the dark hole follows the 88.5\degree N PC (88.8\degree N PG) latitude circle. 

High-resolution morphology of the center of the polar vortex is presented in Fig.~\ref{Fig:NPV_NAC}. The maps show a region within 1.5\degree~of latitude around the north pole. The morphology in CB2 (Fig.~\ref{Fig:NPV_NAC}a), RED (Fig.~\ref{Fig:NPV_NAC}d) and GRN (Fig.~\ref{Fig:NPV_NAC}e) are very similar. The BL2 view (Fig.~\ref{Fig:NPV_WAC}f) exhibits the same dark center as in the context view (Fig.~\ref{Fig:NPV_WAC}g). In BL2, hole lacks a clear outline, and its reflectivity gradually darkens toward the center of the hole. In MT2, the circular dark center (possibly a hole in the upper tropospheric haze) is apparent; this feature can also be seen in the context image (Fig.~\ref{Fig:NPV_WAC}c); inside of the hole, only the brightest of the features that are visible in CB2 (Figs.~\ref{Fig:NPV_NAC}a) are apparent. Figure~\ref{Fig:NPV_NAC}c shows the view in MT3; the circular hole with a well-defined outline that roughly follows the 88.5\degree N PC (88.8\degree PG) latitude is the same feature as that seen in MT2 (Fig.~\ref{Fig:NPV_NAC}b); however, no cloud feature is visible inside of the dark region, indicating a lack of light-scattering aerosols above the upper troposphere. Figure~\ref{Fig:NPV_color} shows the near-infrared false color and an enhanced visible color maps of the NPV.
 
\subsection{North polar wind structure}
\begin{figure*}%
\begin{center}
\figurebox{6.85in}{}{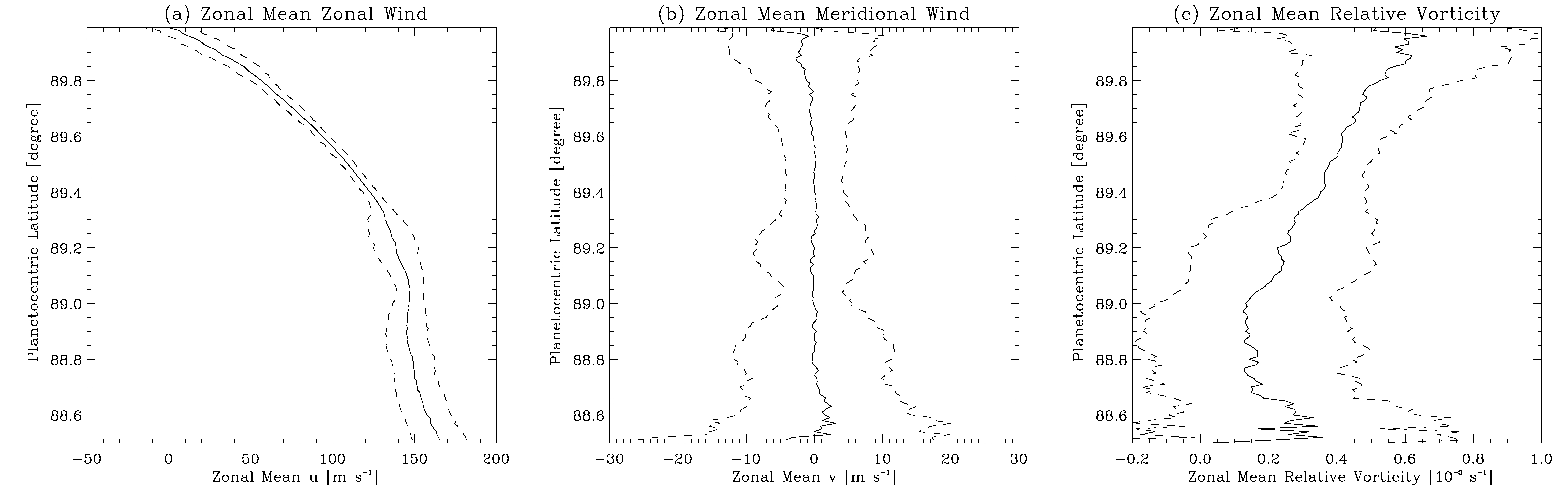}
\caption{The zonal wind structure at the center of the NPV measured using Cassini ISS NAC images. (a) Zonal mean of eastward wind component. (b) Zonal mean of northward wind component. (c) Zonal mean of the relative vorticity. The dashed lines in each panel denote zonal mean plus and minus the zonal standard deviation. From \cite{Sayanagi_etal_2015npv}.}
\label{Fig:NPV_WIND}
\end{center}
\end{figure*}

The cyclonic sign of the NPV's relative vorticity was predicted by the CIRS measurement of the temperature field \citep{Fletcher_etal_2008_CIRS_PoleHex}; however, the region's vorticity was not directly measured until \cite{Baines_etal_2009poles} used VIMS images to analyze the motion of clouds to deduce the zonal-mean wind and vorticity structure in the northern high latitudes. Higher resolution wind measurements have been obtained using ISS images; \cite{Antunano_etal_2015_SatPoleDyn} analyzed the dynamical context surrounding the NPV (Fig.~\ref{Fig:HexZonalWind}E), while \cite{Sayanagi_etal_2015npv} analyzed the high-resolution wind structure at the center of the NPV between the north pole and 88\degree N PC (Fig.~\ref{Fig:NPV_WIND}). 

To the north of the hexagon jet at 75\degree N PC (78\degree N PG), the zonal wind profile has a minimum at around 80\degree N PC (82\degree N PG) latitude \citep{Baines_etal_2009poles, Antunano_etal_2015_SatPoleDyn}. From the minimum, the zonal mean wind speed monotonically increases poleward until reaching the rim of the ``eye'' of the NPV such that the profile has a uniform absolute vorticity (i.e., $\omega_\mathrm{a}=$\emph{constant} where the absolute vorticity is $\omega_\mathrm{a} = f+\zeta$, $\zeta$ is the relative vorticity, $f=2\Omega\sin{\phi}$ is the Coriolis parameter, $\Omega$ is the planetary rotation rate, and $\phi$ is the latitude). The VIMS measurements by \cite{Baines_etal_2009poles} detected a peak wind speed of 136.0$\pm$6.5~m~s$^{-1}$ at 88.3\degree N PC (88.6\degree N PG) latitude, which is near the rim of the NPV's eye. The analysis by \cite{Antunano_etal_2015_SatPoleDyn} using ISS images returned a similar result (Fig.~\ref{Fig:HexZonalWind}).

\adjustfigure{35pt}

\cite{Sayanagi_etal_2015npv} used ISS images to analyze the wind structure at the center of the NPV poleward of 88.5\degree N PC (88.5\degree N PG) latitude. The zonal mean and standard deviation of the eastward wind component $u$ are presented in Figs.~\ref{Fig:NPV_WIND}a. From the north pole, the zonal wind speed steadily increases equatorward to the end of the measurement at 88.5\degree N PC latitude. The zonal mean of the meridional wind component $v$, shown in Fig.~\ref{Fig:NPV_WIND}b, is essentially zero throughout the domain of measurement. Figure~\ref{Fig:NPV_WIND}c presents the result of the zonal relative vorticity around the north pole, showing that the cyclonic relative vorticity peaks at the pole at $\sim$6$\times$10$^{-4}$~s$^{-1}$. This value is slightly greater than that measured by \cite{Baines_etal_2009poles}, $\sim$5$\times$10$^{-4}$~s$^{-1}$, which is not surprising because the ISS measurements give higher spatial resolution and the peak speed at the pole is most likely not resolved in VIMS measurements. The relative vorticity analysis by \cite{Antunano_etal_2015_SatPoleDyn} gives a value of $(2.5\pm0.1)\times10^{-4}$~s$^{-1}$ at 89.8\degree N PC (89.0\degree N PG), which is less than half of that measured by \cite{Sayanagi_etal_2015npv}; this difference is also likely due to the lower-resolution data and spatial averaging employed by \cite{Antunano_etal_2015_SatPoleDyn}.

We note a confusion in the literature regarding the vorticity in the polar regions of Saturn. \cite{Dyudina_etal_2009} used an unconventional definition of vorticity that reversed the sign (defined by the first equation in their Section~4), resulting in a positive sign of vorticity for the cyclonic south polar vortex. \cite{Baines_etal_2009poles} employed the same convention as \cite{Dyudina_etal_2009} to define vorticity, resulting a negative sign of vorticity for the cyclonic north polar vortex. \cite{Antunano_etal_2015_SatPoleDyn} employs a conventional definition of relative vorticity (their Eq.~4); however, their calculation contains a sign error, resulting in a negative sign of relative vorticity for the cyclonic north polar vortex. In this review, we compare the measurements by these authors using the conventional definition of the relative vorticity, i.e., a cyclonic vortex has a positive relative vorticity in the northern hemisphere, and negative in the southern hemisphere.

\section{Comparison of the northern high latitudes to the south polar region}
\label{section:n-s-compare}

In this section, we compare the cloud morphology of the north-polar vortex to that of the south-polar vortex. The morphology and dynamics of Saturn's south pole has been documented by \cite{Sanchez-Lavega_etal_2006}, \cite{Dyudina_etal_2008, Dyudina_etal_2009}, \cite{Antunano_etal_2015_SatPoleDyn} and \cite{Sayanagi_etal_2015npv}. The existence of a cyclonic vortex at the south pole was first indicated by a hotspot found in earth-based observations by \cite{Orton_Yanamandra-Fisher_2005} and later in a CIRS-derived temperature map at the 100-mbar level by \cite{Fletcher_etal_2008_CIRS_PoleHex}. Cassini imaging observations \citep{Vasavada_etal_2006, Sanchez-Lavega_etal_2006} revealed cyclonic rotation around the spot in 2005. 

\begin{figure}%
 \begin{center}
 \figurebox{3.4in}{}{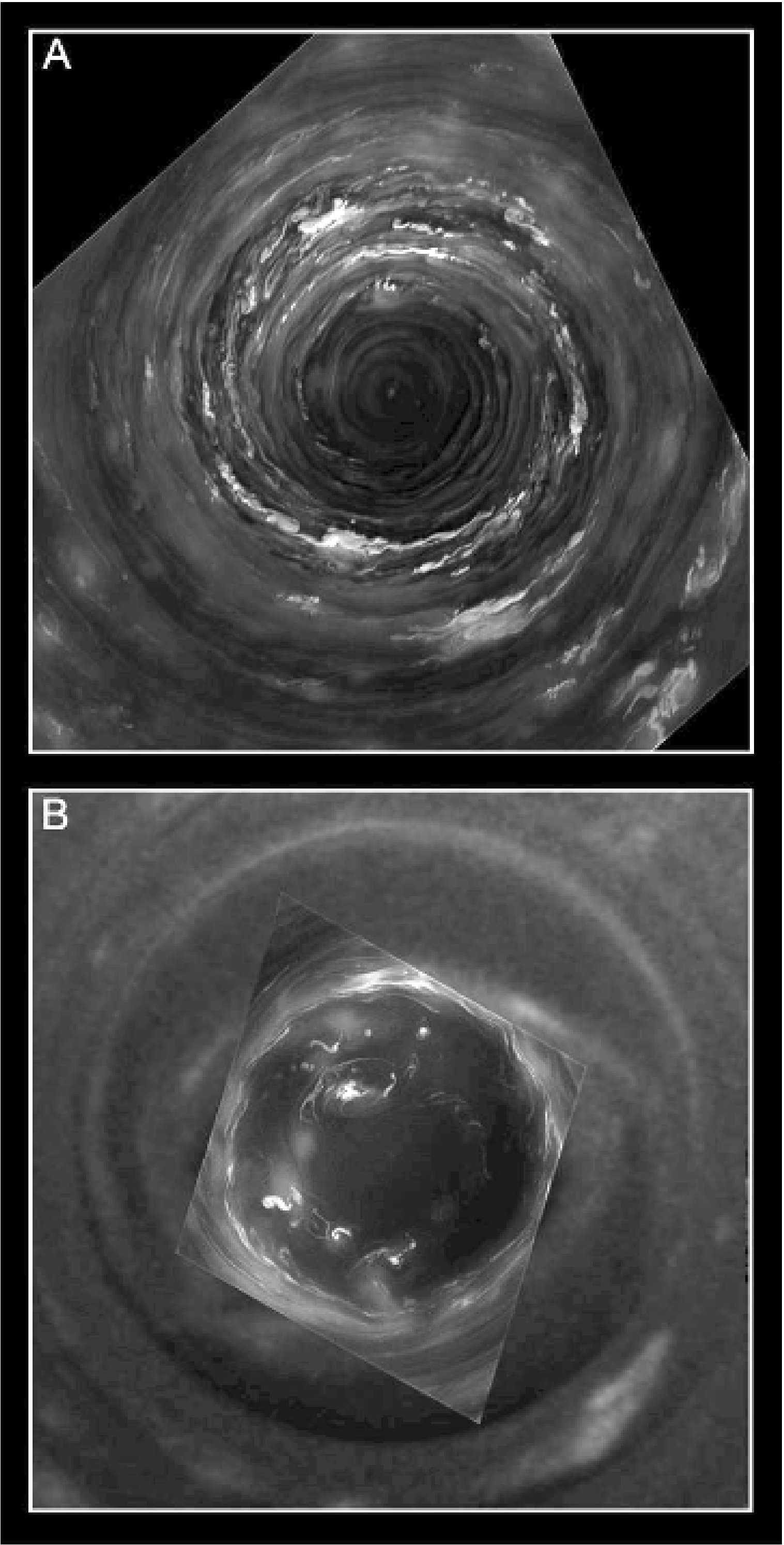}
 \caption{(a) Polar projection of the north polar region from 87.5 to 90 degrees latitude on 14 June 2013, using an image with CB2 filter and spatial resolution at the pole of 5.3 km/pixel. (b) Polar projection of the south polar region from 87.5 to 90 degrees on 14 July 2008. Panel b is a composite of two different images with different resolutions. The image on the front is a NAC image with CB2 filter at 2.4 km/pixel, while the one at the back is a WAC image with a CB2 filter at 29.5 km/pixel. From \cite{Antunano_etal_2015_SatPoleDyn}.}
\label{Fig:North_South_Compare}
 \end{center}
 \end{figure}

Figure~\ref{Fig:HexMorph} panels A and B (from \citealt{Antunano_etal_2015_SatPoleDyn}) compare large-scale structure of north and south polar regions poleward from 60\degree north and south latitude. The polar vortex at the south pole appears brighter, but this is an artifact of the mosaicking and overcorrection at the latitudes where the illumination incident angle is nearly 90\degree. Without limb-darkening correction, the south polar vortex has a contrast similar to that of the northern vortex, as can be seen in Figure \ref{Fig:North_South_Compare}, which shows a high-resolution polar projection of both poles of Saturn at the same scale.

\adjustfigure{40pt}

 \begin{figure}%
 \begin{center}
 \figurebox{3.4in}{}{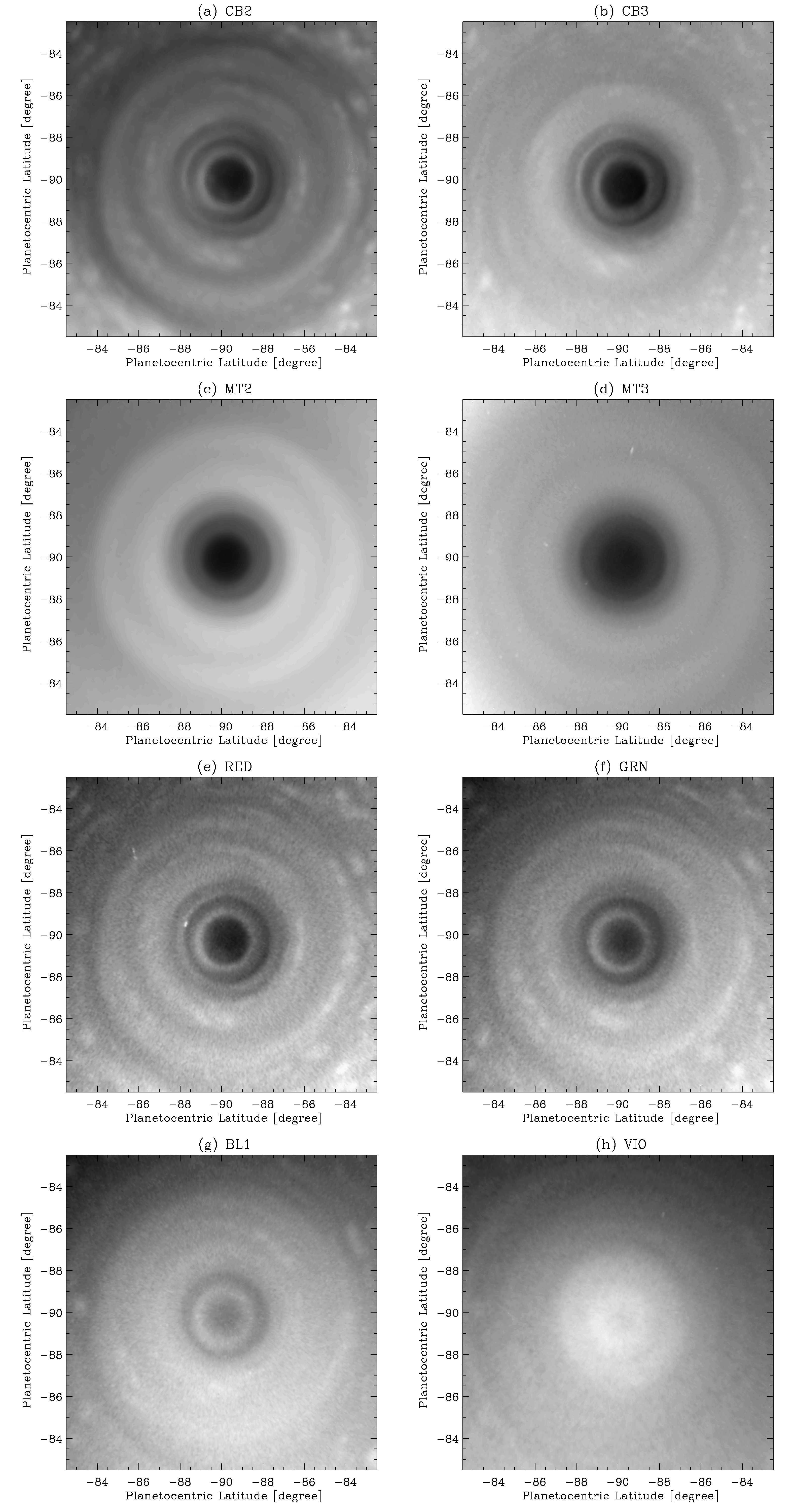}
\caption{Same as Fig.~\ref{Fig:NPV_WAC}, except that south polar data captured in 2007 is shown here. From \cite{Sayanagi_etal_2015npv}.}
\label{Fig:SouthPole}
 \end{center}
 \end{figure}

Figure~\ref{Fig:SouthPole} presents polar-projected maps of the southern high-latitude region using various filters. In CB2 and CB3 images (Figs.~\ref{Fig:SouthPole}a and \ref{Fig:SouthPole}b), the SPV exhibit cyclonically spiraling morphology centered on the pole and extending down to around 83\degree S PC (84\degree S PG) latitude. The filamentary clouds in the images demonstrate wind shear in the polar vortices. The cloud texture seems richer and more complex around the north pole, which contains more fine-scale structures. The difference is primarily due to the difference in the high-altitude haze, although effects of the observation geometry are also not negligible. The eastward wind speed peaks at the edge of the ``eye'' of the cyclonic vortex in both the NPV and the SPV. For both NPV (see Fig.~\ref{Fig:NPV_WAC}) and SPV (Fig.~\ref{Fig:North_South_Compare}), the edge of the eye appears as rings of bright clouds. The peak wind speeds at 88\degree-89\degree north and south latitudes reach about 150~m~s$^{-1}$ as shown in Fig.~\ref{Fig:HexZonalWind} panels E and F, and Fig.~\ref{Fig:NPV_WIND}a. 

The double-wall structure surrounding the ``eye'' of the south polar vortex, originally reported by \cite{Dyudina_etal_2008}, is apparent when viewed in CB2 (Fig.~\ref{Fig:SouthPole}a), CB3 (\ref{Fig:SouthPole}b), RED (\ref{Fig:SouthPole}e), GRN (\ref{Fig:SouthPole}f) and BL1 (\ref{Fig:SouthPole}g). In those filters, the interior of the eye appears dark. Outside of the eye, a concentric albedo structure dominates the morphology, interior of which appears substantially darker than the outside. A few diffuse discrete features can also be seen inside of the eye. The interior of the eye also appears dark in MT2 (Fig.~\ref{Fig:SouthPole}c) and MT3 (\ref{Fig:SouthPole}d). While the interior of the eye of the south-polar vortex appears darker than the surroundings in all above filters, the contrast is reversed in VIO (\ref{Fig:SouthPole}h), in which a bright hood covers the south-polar vortex eye region poleward of 87\degree S PC latitude; the polar hood is apparent only in the VIO filter because it is sensitive to particulate scattering in the stratosphere. Radiative transfer analyses are needed to deduce the altitudes of these cloud and haze layers, and to infer the properties of the aerosol particles that make up those clouds and hazes; such analyses are yet to be done.

\begin{figure*}%
\begin{center}
\figurebox{6.85in}{}{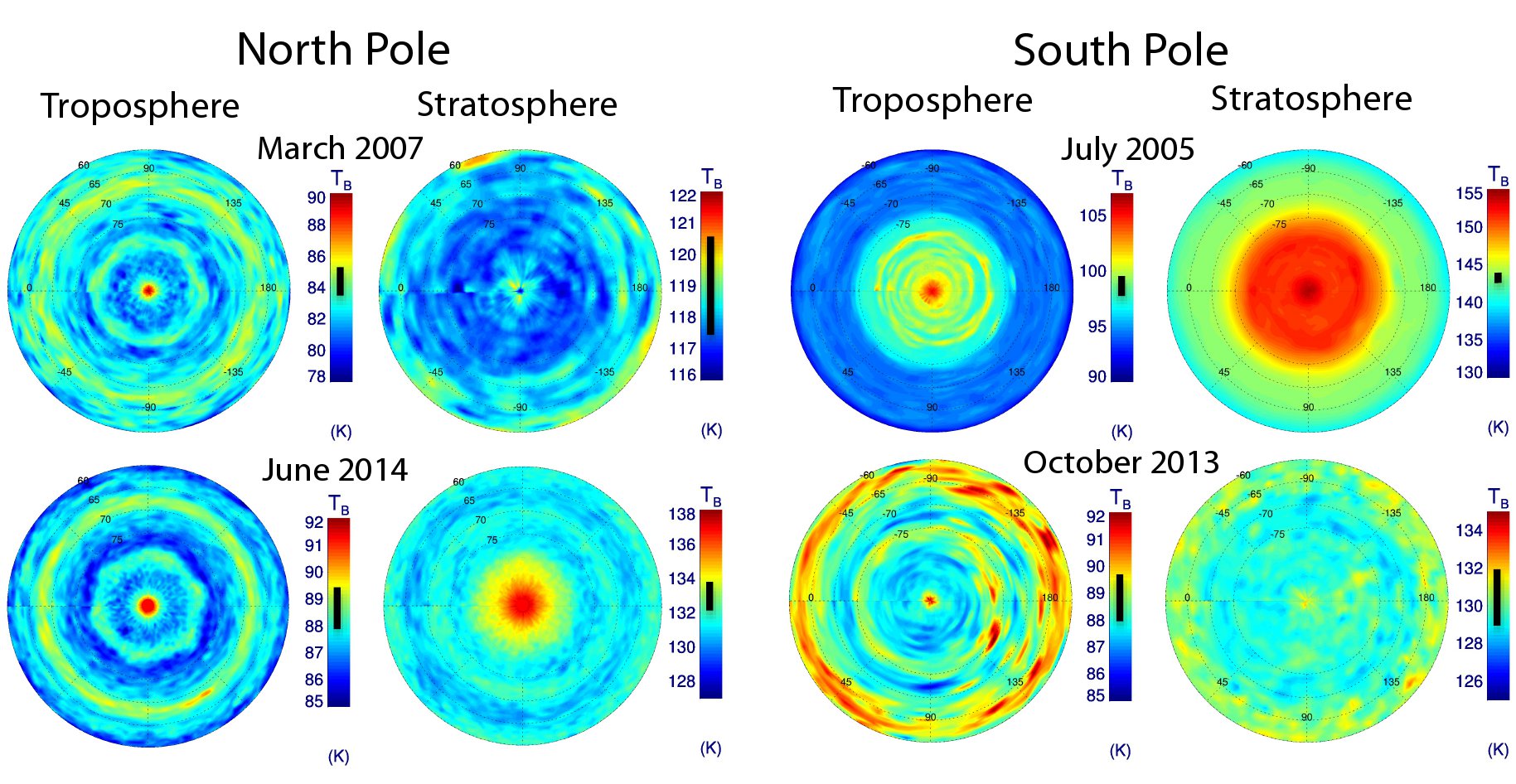}
\caption{Brightness temperature maps of the troposphere and stratosphere constructed from Cassini/CIRS spectroscopy. Tropospheric maps use the H$_2$-He collision-induced continuum between 600-620 cm$^{-1}$ to sense the 150-mbar level; stratospheric maps use methane emission between 1290-1310 cm$^{-1}$ to sense the 1-mbar level. Black lines in each scale bar indicate the uncertainty on the measurements, which . The top row shows the polar regions in northern winter \citep[adapted from][]{Fletcher_etal_2008_CIRS_PoleHex}, the bottom row shows the polar regions in northern spring \citep[adapted from][]{Fletcher_etal_2015SaturnPoles}.}
\label{Fig:polarmaps}
\end{center}
\end{figure*}
  
A cyclonic vortex with a compact region of enhanced thermal infrared emissions centered on each of the poles has been a persistent feature of the polar tropospheres for the full ten-year span of the observations in cloud-tracking wind measurements as well as thermal fields, despite seasonal shifts in the absolute temperatures (Fig. \ref{Fig:polarmaps}). The warm air is isolated from lower latitudes (cool tropospheric polar zones near $\pm$84\degree PC (85\degree PG) latitude) by peripheral jets at 88.6\degree N PC (88.9\degree N PG) and 87.5\degree S PC (88.0\degree S PG), respectively. Warm tropospheric belts exist equatorward of the polar zones, and poleward of the eastward jets at 75.4\degree N PC (78.0\degree N PG; i.e., the hexagon jet) and 70.5\degree S PC (73.9\degree S PG) latitudes. The visibility of the south polar belt has increased as the upper tropospheric temperature gradients have changed, and the warm north polar belt has maintained its hexagonal wave structure throughout the mission \citep{Fletcher_etal_2015SaturnPoles}. Because the northern jet's peak at 75.4\degree N PC (78.0\degree N PG) is located at a higher latitude than the southern jet at 70.5\degree S PC (73.9\degree S PG) as can be seen in Figs.~\ref{Fig:HexZonalWind}E and \ref{Fig:HexZonalWind}F, the morphologies of the northern and southern polar belts are fundamentally different --- the northern belt sits closer to the pole than the south, confined by the eastward jets.

\section{Polar stratosphere}
\label{section:strat}
This section discusses the thermal structure and hydrocarbon distribution in the polar regions as well as their response to seasonal radiative forcing. Atmospheric temperatures are determined by a complex interplay between radiative energy balance and atmospheric circulation, which is in turn determined by the contributions of composition and aerosols to the radiative heating and cooling (see Chapter~10). By deriving the thermal, chemical and aerosol distributions within and surrounding the polar vortices, we are able to gain insights into the circulations shaping these unique environments on Saturn from the upper troposphere into the stratosphere. Vertical motions are inferred from both the temperature field (via the thermodynamic energy relation) and the use of gaseous species as quasi-conservative tracers (via the continuity equation). Horizontal motions are determined from latitudinal temperature gradients via the thermal wind equation, assuming that zonal winds are in geostrophic balance. 

\subsection{Pre-Cassini observations}
Prior to Cassini's arrival in 2004 (planetocentric solar longitude of $L_\mathrm{s}=293^\circ$), our knowledge of the polar stratosphere was restricted to the Earth-facing summer hemisphere. \citet{Orton_Yanamandra-Fisher_2005} discovered a region of elevated south polar stratospheric methane emission in Keck observations in February 2004 ($L_\mathrm{s}=287^\circ$, just after southern summer solstice), poleward of approximately 65\degree S PC (70\degree S PG) and sensitive to the 0.5-20~mbar pressure range. This mirrored a similar region of enhanced emission observed in March~1989 in the northern hemisphere using the Infrared Telescope Facility \citep[$L_\mathrm{s}=104.5^\circ$, just after northern summer solstice,][]{Gezari_etal_1989_Sat-IR}, and suggested the presence of a seasonally-varying stratospheric vortex poleward of approximately 65\degree PC (70\degree PG) latitude in both hemispheres. The ground-based record has insufficient temporal sampling to constrain the timescales for the formation and dissipation of these summertime vortices, presenting a unique opportunity for Cassini. 

\adjustfigure{85pt}

\begin{figure*}%
\begin{center}
\figurebox{6.85in}{}{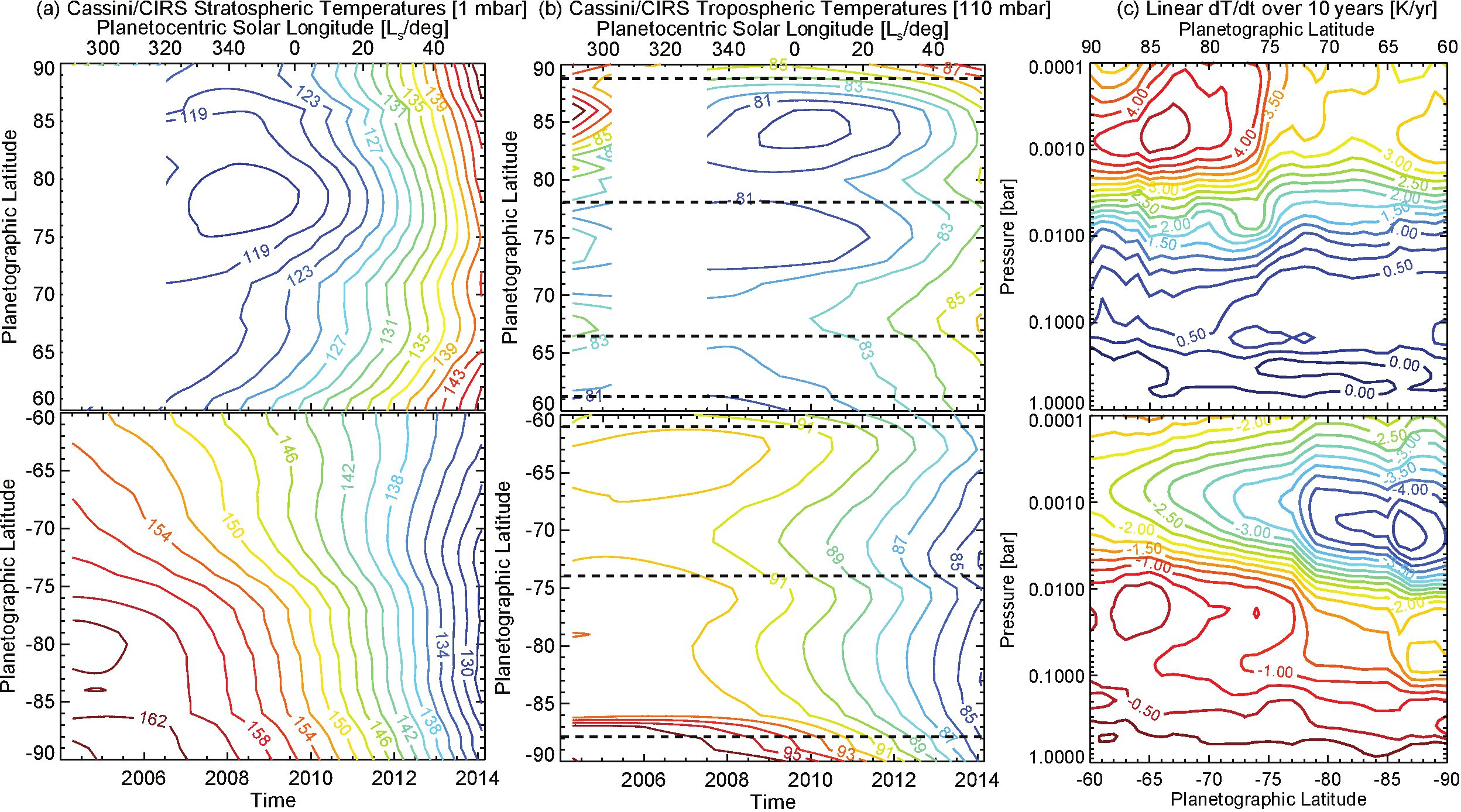}
\caption{Measured temperature changes over the ten years of the Cassini mission for the stratosphere (1~mbar, column a) and troposphere (110~mbar, column b), reconstructed from Cassini/CIRS measurements \citep{Fletcher_etal_2015SaturnPoles}. In column b the horizontal dashed lines indicate the peaks of the eastward tropospheric jets to show the correspondence with the temperature field. Column c provides a simple way of visualizing the temperature changes $\mathrm{d}T/\mathrm{d}t$ [K/yr] assuming a linear change. Although this is evidently not the case in columns a and b, it serves to demonstrate that the most intense heating and cooling is occurring within $\pm15^\circ$ of the poles.}
\label{Fig:Tchanges}
\end{center}
\end{figure*}

In addition to the large region of enhanced methane emission poleward of $\sim$65\degree S PC ($\sim$70\degree S PG) latitude, \citet{Orton_Yanamandra-Fisher_2005} also detected a warm, compact cyclonic vortex in the summer stratosphere and upper troposphere, centered on the south pole with a radius of 2-3\degree~latitude. This warm cyclone was spatially coincident with the hurricane-like eyewalls identified in the tropospheric cloud decks in visible light \citep{Sanchez-Lavega_etal_2006, Dyudina_etal_2008, Dyudina_etal_2009, Baines_etal_2009poles, Antunano_etal_2015_SatPoleDyn, Sayanagi_etal_2015npv}, suggesting that the circulation responsible for the cloud-deck vortex was penetrating through the stably-stratified upper troposphere and high into the stratosphere over the summer pole. Later measurements by Cassini/CIRS \citep{Fletcher_etal_2008_CIRS_PoleHex} measured 1-mbar temperatures in excess of 165 K within the compact south polar cyclone. 

\subsection{Thermal structure of the north polar vortex}
Cassini's arrival at Saturn in 2004 allowed characterization of the thermal structure of the north pole, which had been hidden from Earth-based observers since the mid-1990s. \citet{Fletcher_etal_2008_CIRS_PoleHex} utilized Cassini/CIRS observations to compare the summer and winter poles using data acquired between 2005 and 2007 ($L_\mathrm{s}=300-340^\circ$), and discovered a compact spot of enhanced thermal emission in the troposphere of the northern winter pole, mirroring that observed at the south pole. As mentioned in Section~\ref{section:npv}, the north polar hotspot was later shown to be coincident with the north polar vortex by the motion of the clouds observed in 5-$\mu$m by \cite{Baines_etal_2009poles}. However, unlike the summertime south polar cyclone, there was no evidence of the compact vortex penetrating high into the wintertime northern polar stratosphere (see North Pole in March~2007 in Fig.~\ref{Fig:polarmaps}). As discussed in Section~\ref{section:hex}, the north polar belt near 70-78\degree N PC (75-80\degree N PG) exhibited the sinusoidal manifestation of the hexagon in the upper troposphere that did not penetrate into the overlying stratosphere. However, contrary to the expectations of radiative climate models that predict the coldest stratospheric temperatures to occur right at the north pole (see Chapter~10), \citet{Fletcher_etal_2008_CIRS_PoleHex} demonstrated that the high northern latitudes were subtly warmer than lower latitudes, presumably as a result of adiabatic heating caused by large-scale atmospheric subsidence.

As of this writing, the temperature and composition of the northern and southern polar regions (upper troposphere and stratosphere) have been tracked by Cassini from northern winter ($L_\mathrm{s}=293^\circ$) through to northern spring ($L_\mathrm{s}=56^\circ$), using deviations from radiative-climate and chemical expectations to infer the general circulation of Saturn's polar stratosphere \citep{Fletcher_etal_2015SaturnPoles}. 

\adjustfigure{90pt}

\subsection{Seasonal variation}
\label{section:seasonal_variation}
Above the tropopause, radiative climate models that balance radiative heating via methane and aerosol absorption with radiative cooling via hydrocarbons (\citealt{Bezard_Gautier_1985_GP-Climes, Greathouse_etal_2010_SatSeasonalTemps, Friedson_Moses_2012, Guerlet_etal_2014_SaturnGCM}) predict the rate of change of temperatures $\mathrm{d}T/\mathrm{d}t$ to vary monotonically with latitude such that the coldest winter temperatures would be expected right at the poles. Given that the atmosphere has a substantial thermal inertia that increases with depth, these coolest stratospheric temperatures lag behind the winter solstice by approximately a season. Indeed, Cassini detected the coldest northern temperatures between 2008-2010, 6-10 years after the winter solstice (\citealt{Fletcher_etal_2015SaturnPoles}; Fig.~\ref{Fig:Tchanges}). Peak stratospheric temperatures in the southern hemisphere occurred 1-2 years after southern summer solstice, consistent with radiative climate predictions. However, the \textit{location} of the coolest northern temperatures was at 72-78\degree N PC (75-80\degree N PG), with the stratosphere warming at higher latitudes. Indeed, the northern stratosphere within $15^\circ$ latitude of the pole has been warming throughout the Cassini observations, by approximately 5 K/year at 0.5-1.0 mbar. This has been mirrored at the south pole by a cooling within $15^\circ$ of the pole at a similar rate, but with the peak cooling occurring at slightly higher pressures of 1-3 mbar. A crucial observational difference is that a strong north-south temperature gradient at the 1-mbar level occurred at 72\degree S PC (75\degree S PG) during southern summer, giving the appearance of an isolated stratospheric vortex, whereas no such gradients are yet visible in the northern stratosphere as of the time of writing (Spring 2015). Based on the observations of \citet{Gezari_etal_1989_Sat-IR} shortly after the previous northern summer solstice, we can expect that such a sharp boundary will form near 72\degree N PC (75\degree N PG) between now and the end of the Cassini mission ($L_\mathrm{s}=93^\circ$), signalling the onset of the summertime vortex. The 72\degree S PC (75\degree S PG) boundary vanished after the autumnal equinox as the southern vortex cooled by more than 35~K, implying the dissipation of the southern summer stratospheric vortex.

\subsection{Implications for zonal wind vertical shear}
The shifting latitudinal temperature gradients in the upper troposphere and stratosphere discussed in Section~\ref{section:seasonal_variation} have implications for the shear on the zonal jets measured at the tropospheric cloud level. For example, Fig.~\ref{Fig:Tchanges} column (a) shows that the jet at 70.5\degree S PC (73.9\degree S PG) (see Fig.~\ref{Fig:HexZonalWind}) exists in a region of temporally evolving horizontal temperature gradient; the thermal wind equation (see, e.g., \citealt{Holton_2004}) can be used to calculate the vertical shear and extrapolate the vertical wind profile from the tropospheric wind measurements like those shown in Fig.~\ref{Fig:HexZonalWind}. The change in the horizontal temperature gradient indicates that the stratospheric wind speed at 1~mbar was westward at the start of the Cassini mission, but weakened and became eastward between 2004 and 2014. In comparison, the horizontal temperature gradient around the jet at 75.4\degree N PC (78.0\degree N PG) has not evolved substantially in that time period. The overlying stratospheric trends promote positive vertical shear in the winter hemisphere and negative vertical shear in the summer hemisphere \citep{Friedson_Moses_2012}. Similar oscillations in the stratospheric jet velocities were observed for \emph{some} high-latitude jets in the radiative dynamical model of \citet{Friedson_Moses_2012}, but not for all. However, calculating the vertical shear by integrating the thermal wind equation is extremely sensitive to uncertainties the horizontal temperature gradient, and the inferred vertical shear remains to be tested via direct measurements; remote-sensing wind measurements at the 1-mbar level is difficult due to the absence of observable cloud tracers in the stratosphere. Changes in jet velocities in the upper troposphere are expected to be smaller than 10~m~s$^{-1}$, and are likely to be undetectable in cloud-tracking observations in the main cloud deck. The implications of the thermal wind equation are that the strong negative $\mathrm{d}T/\mathrm{d}y$ at 72\degree S PC (75\degree S PG) in southern summer is related to a westward jet encircling the region of warm emission at the south pole. No similar westward jet is yet present at 72\degree N PC (75\degree N PG). 

\begin{figure}%
\begin{center}
\figurebox{3.6in}{}{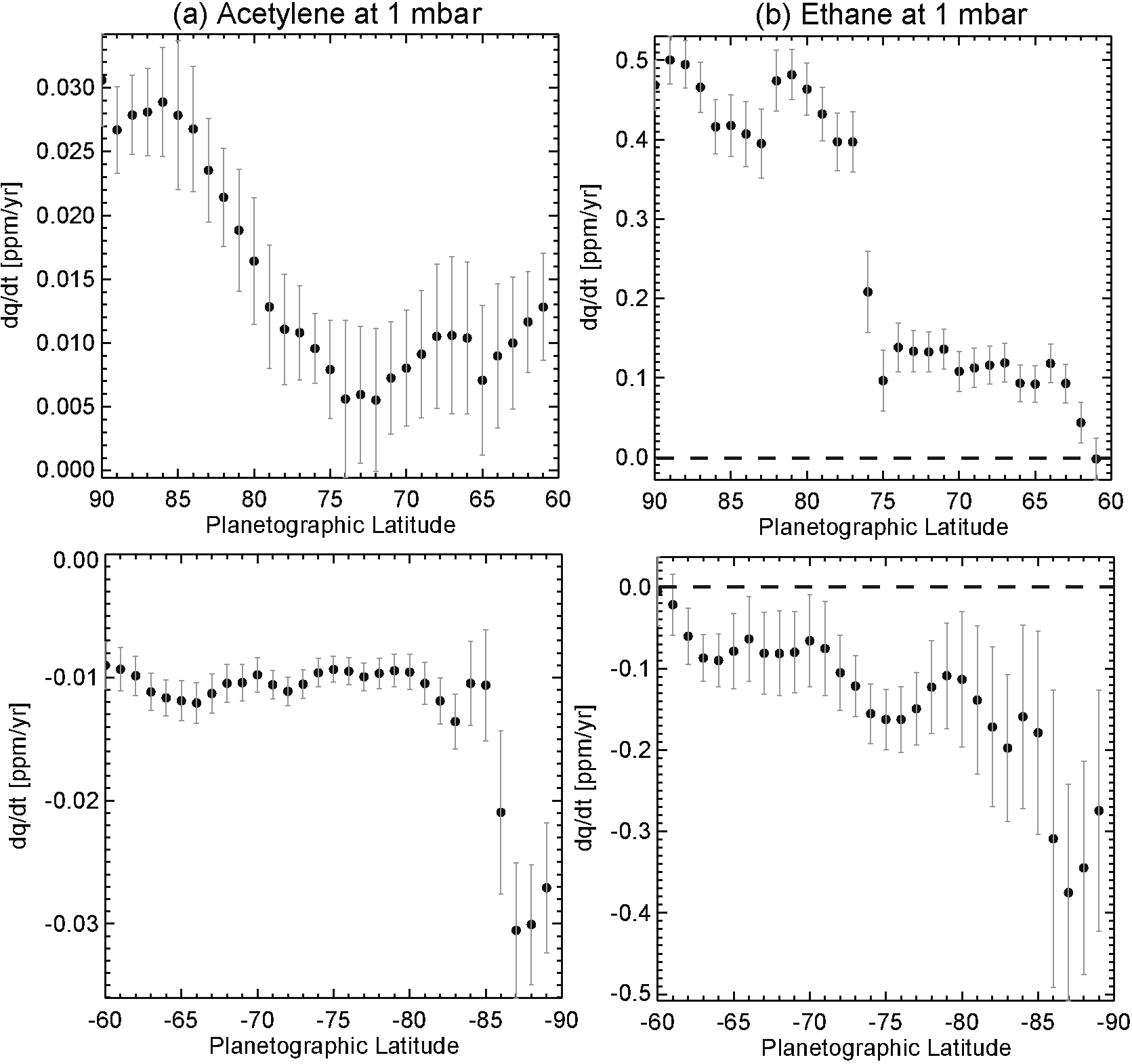}
\caption{Rate of change of mole fraction ($q$) with time [ppm/year] at the 1-mbar level as measured by Cassini/CIRS spectroscopy, assuming a linear dependence over the ten-year span of the Cassini mission. Acetylene and ethane respond to the seasonal changes in insolation in subtly different ways, but are consistent with subsidence enhancing hydrocarbons over the north polar region, and upwelling depleting hydrocarbons over the south polar region. From \citet{Fletcher_etal_2015SaturnPoles}.}
\label{Fig:polarcxhy}
\end{center}
\end{figure}

\subsection{Effects on the hydrocarbon distribution}
In both hemispheres of Saturn, the polar-most zonal jets may act as a transport barrier that isolates the unique chemistry and aerosols of the polar regions from the lower and mid-latitudes. Over much of Saturn's stratosphere, neutral photochemistry dominates the sources and sinks of the soup of hydrocarbon species produced by the UV photolysis of methane \citep[e.g.,][]{Moses_Greathouse_2005}. These hydrocarbon products sediment downwards to ultimately be recycled in the troposphere, but some can condense to form high altitude stratospheric haze layers that are typically optically thin to remote sensing, but can be seen in limb views from Cassini \citep[e.g.,][]{Rages_Barth_2012}. At the poles, however, energy inputs associated with auroral energy and heating, in addition to ion chemistry, have the potential to shift the atmospheric chemistry away from the expectations of neutral photochemistry. Numerical simulations of ion chemistry are yet to fully explain observed aerosol properties \citep{Friedson_etal_2002JupPolarHaze, Wong_etal_2000JupPolarHaze, Wong_etal_2003JupPolarHaze}, so we still rely on photochemistry to inform our expectations for the polar composition, which suggests that the 1-mbar abundances of ethane and acetylene should not be varying significantly with time (although lower $\mu$bar pressures will see relatively rapid increases in hydrocarbon production when sunlight returns to the high latitudes in the spring, peaking at solstice and then declining through autumn). \cite{Guerlet_etal_2015_SatAerosols}'s thermal infrared detection of polar aerosols provides observational evidence that the auroral chemistry leads to the production of these aerosols.

Cassini has captured snapshots of the zonal mean hydrocarbon distributions during southern summertime conditions (see Chapter~10), finding considerable asymmetries between northern and southern hemispheres. Measurements at high latitudes have been sparse. \citet{Sinclair_etal_2013} identified a minimum in the 2-mbar acetylene (C$_2$H$_2$) distribution near 78\degree S PC (80\degree S PG), followed by a sharp rise at higher latitudes within the summer stratospheric vortex. Similar, shallower latitudinal gradients were also observed in ethane towards the south pole and acetylene towards the north pole. Higher order hydrocarbons might also be expected to be enhanced at high polar latitudes, such as the recent detection of benzene (potentially related to ion chemistry) at Saturn's south pole \citep{Guerlet_etal_2015_SatAerosols}. \citet{Fletcher_etal_2015SaturnPoles} report on the rate of change of the C$_2$H$_6$ and C$_2$H$_2$ abundances at both poles over ten years of Cassini observations (Fig.~\ref{Fig:polarcxhy}). Hydrocarbon emissions were barely discernible at the north pole during winter because of the low temperatures, and although the abundances of both C$_2$H$_6$ and C$_2$H$_2$ have been steadily rising poleward of 72\degree N PC (75\degree N PG) during spring (0.45~ppm~year$^{-1}$) for ethane and 0.03~ppm~year$^{-1}$ for acetylene), there is no peak in the hydrocarbon abundances within $\pm5^\circ$ of the north pole as there appeared to be in the south. Poleward of 72\degree S PC (75\degree S PG), the abundances have been steadily falling, by 0.35~ppm~year$^{-1}$ for ethane and 0.03~ppm~year$^{-1}$ for acetylene. As these rates far exceed the expectations of diffusive photochemistry models, we invoke subsidence of hydrocarbon-rich air at the north pole, and upwelling of hydrocarbon-poor air at the south pole, to explain these observed changes to the hydrocarbon distributions. 
\subsection{Vertical motion}
In a planetary stratosphere, the vertical component of the general circulation is controlled by the radiative heating and cooling. Even though the speed of the vertical motion is too slow for direct measurement, it can be estimated from the deviation of the temperature from the radiative equilibrium. Thus, the calculation of radiation energy balance is the first step in understanding the stratospheric dynamics. The key radiative coolants in Saturn's stratosphere are ethane and acetylene; however, the current generation of radiative-dynamical models (i) lack the spatial resolution to deal with the hydrocarbon excesses within the polar regions (and the enhanced cooling they may produce), (ii) hold the hydrocarbons fixed with time rather than allowing them to vary seasonally; and (iii) do not permit the advection of these hydrocarbons as quasi-conserved tracers. Furthermore, polar stratospheric aerosols are expected to provide an additional term to the radiative budget \citep{Fletcher_etal_2015SaturnPoles, Guerlet_etal_2015_SatAerosols}, enhancing the rate of summertime heating and wintertime cooling. To address these issues, ongoing efforts focus on more accurate modeling of the radiative balance in Saturn's stratosphere.

Although the current models are incomplete, deviations from expected radiative equilibrium can be assessed in terms of vertical motions causing adiabatic heating and cooling \citep[e.g.,][]{Conrath_etal_1990}. In this sense, the warming north polar temperatures, which increased at their fastest rates poleward of 72\degree N PC (75\degree N PG), could be related to subsiding air masses at vertical velocities of $w\approx-0.1$~mm~s$^{-1}$ \citep{Fletcher_etal_2015SaturnPoles}. Similarly, the cooling of the southern polar vortex could be related to upwelling at $w\approx+0.1$~mm~s$^{-1}$. These estimates, derived from a modified form of the heat equation \citep{Conrath_etal_1990}, are only approximate given the large uncertainties in the absolute values of the radiative equilibrium temperature. Nevertheless, vertical motions of similar magnitude can explain the observed variability of the hydrocarbon distributions, and suggest that springtime subsidence and autumnal upwelling are indeed occurring, and that these are related to the formation/dissipation of the northern/southern summertime stratospheric vortices.

This stratospheric circulation, potentially confined to latitudes poleward of $\pm$72 \degree PC (75\degree PG) in both hemispheres, may have consequences for the distributions of clouds, hazes, and chemicals throughout the polar regions. For example, stratospheric polar hazes are known to change character within the polar regions \citep{Sanchez-Lavega_etal_2006}, being composed of smaller (0.1 $\mu$m radius at high latitude compared to 0.2 $\mu$m at low latitudes, \citealt{Perez-Hoyos_etal_2005}), more UV absorbent (\citealt{Karkoschka_Tomasko_1993}; also see Section~\ref{section:aerosols} and Fig.~\ref{Fig:UVIShex}) and with an increased optical depth \citep{Karkoschka_Tomasko_2005}. In the upper troposphere, the disequilibrium species PH$_3$ is known to be suppressed within the core of the two matching polar cyclones, potentially due to atmospheric subsidence within the centers of the two vortices \citep{Fletcher_etal_2008_CIRS_PoleHex}. The connection between the observed gaseous and aerosol properties and the residual mean circulation of the stratosphere remains unclear. 

In summary, stratospheric and upper-tropospheric circulation can be inferred from measurements of the temporal variability of the thermal structure and the distribution of chemicals (notably stratospheric C$_2$H$_2$ and C$_2$H$_6$) as quasi-conserved tracers. Since Cassini's arrival at Saturn, a region within $15-20^\circ$ (14,000$-$19,000~km) of the south pole, accompanied by bright infrared emission, has been cooling with time (compare the July~2005 and October~2013 maps in Fig.~\ref{Fig:polarmaps}) with declining hydrocarbon abundances, whereas the northern springtime pole has been steadily warming (compare the March~2007 and June~2014 panels in Fig.~\ref{Fig:polarmaps}) with rising hydrocarbon abundances. At the time of writing ($L_\mathrm{s}\approx60^\circ$), the north polar stratosphere lacks a strong meridional temperature gradient near 72\degree N PC (75\degree N PG) that would signify the establishment of a summertime vortex, which we expect to occur before northern summer solstice in 2017. In addition to the seasonally-variable stratospheres, both poles feature warm cyclonic vortices in their upper tropospheres, which extend well into the stratosphere during summertime conditions. 

\subsection{Long-term prediction}
Finally, the success of radiative-climate modeling in reproducing the general thermal trends allows us to make some predictions. Northern summer solstice will be quickly followed by peak 1-mbar temperatures at the north pole (1-2 years after solstice, around 2018-2019) and the presence of a warm northern stratospheric vortex, but given that northern summer solstice occurs near aphelion the magnitude of the temperatures will be smaller than those observed during southern summer (near perihelion). The south pole is currently descending into winter, but we would expect the coolest 1-mbar stratospheric temperatures to occur near to the next equinox (i.e., ring-plane crossing) in 2025. By then, the subsidence at the north pole may cease, the subsidence at the south pole may start, the northern summer vortex would dissipate, and we would start to observe renewed heating of the south pole at 1~mbar for the creation of a new, seasonal, south polar stratospheric vortex.

\begin{figure}%
\begin{center}
\figurebox{3.4in}{}{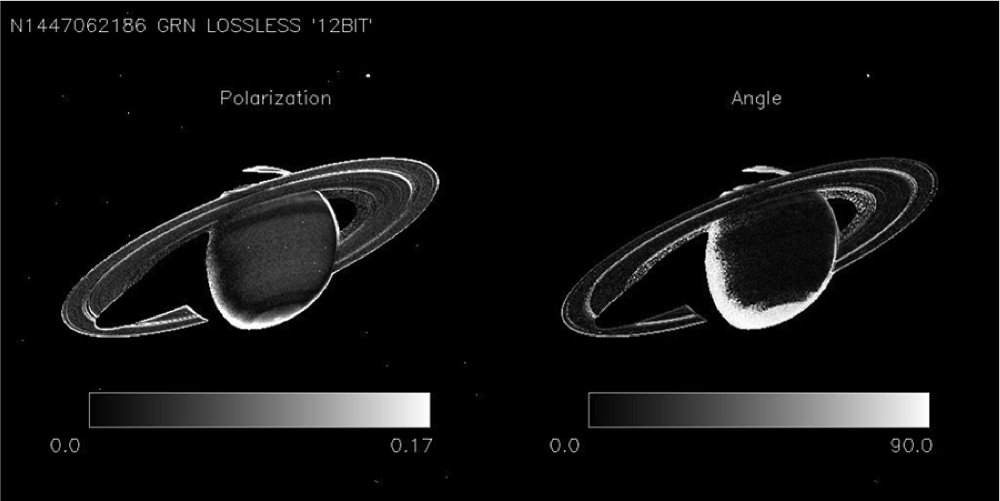}
\caption{Polar polarization measurements, The degree of linear polarization (left) and angle of polarization (right) with respect to the scattering plane captured in 2003 by the Cassini ISS NAC using the GRN filter are shown, depicting enhanced positive polarization (angle of the electric vector near 90\degree with respect to the scattering plane) at high southern latitudes. (From \citealt{West_etal_2009SatBook})}
\label{Fig:Polarization}
\end{center}
\end{figure}

\section{Saturn's polar aerosols and clouds}
\label{section:aerosols}
Polar latitudes are of particular interest for clouds and aerosols as they are for atmospheric dynamics and chemistry. Both Jupiter and Saturn show enhanced ultraviolet absorption at high latitudes and there is strong evidence that energy deposition from the aurora acting on methane contributes to the production of UV absorbers and aerosols in the auroral deposition region (high atmosphere). These effects are more subtle on Saturn than on Jupiter because the auroral energetics are weaker on Saturn. For an introduction to the relevant chemistry, see papers by \cite{Friedson_etal_2002JupPolarHaze} and \cite{Wong_etal_2000JupPolarHaze, Wong_etal_2003JupPolarHaze}. At the same time, images at mid-visible to near-infrared continuum wavelengths show abundant small-scale convective clouds and vortices in the deeper (pressure of $\sim$2 bars or more) atmosphere. The challenges are to sort out the various contributions from this vast stratigraphy, to understand how they influence data at different wavelengths, polarization states, viewing geometry, and time, and then to properly attribute the observed features to atmospheric processes. This effort is still ongoing.

\begin{figure}%
\begin{center}
\figurebox{3.4in}{}{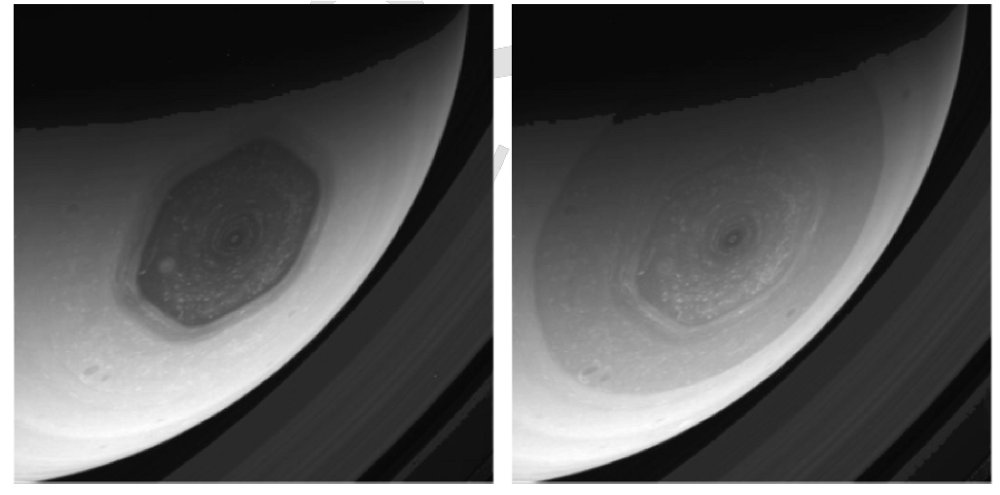}
\caption{Cassini ISS images of Saturn taken 90~s apart in the 935-nm filter, coupled with polarizers with the electric vector parallel (left) and perpendicular (right) to the scattering plane. These images capture Saturn's northern polar region including the hexagon and polar vortex. Structural differences inside and outside the hexagon are apparent. (Figure from \citealt{West_etal_2015PolarimetryBook})}
\label{Fig_HexPolarization}
\end{center}
\end{figure}

Evidence for a population of fractal aggregate particles in the polar stratosphere comes from a combination of images in near-infrared methane bands, images in the near-UV, and polarization images from the Cassini ISS (Imaging Science Subsystem) cameras. In these images, it is clear that the light scattered by the stratospheric haze particles deviate significantly from the scattering phase function predicted by the Mie theory that assumes spherical particles. The polarization signatures suggestive of aspherical aerosols for the southern and northern high latitudes are evident in Fig.~\ref{Fig:Polarization} and Fig.~\ref{Fig_HexPolarization}. However, to date, few analyses have taken advantage of these polarization signatures, and quantitative analysis of light scattering by these polar clouds and hazes is yet to be done --- studies to date have focused on lower latitudes assuming spherical particles (e.g., \citealt{Roman_etal_2013}).

In Fig.~\ref{Fig:Polarization}, taken on Cassini's approach to Saturn, the degree of linear polarization increases at the high latitudes and the angle of polarization at those latitudes is 90 degrees (perpendicular to the scattering plane; \citealt{West_etal_2009SatBook}). These attributes are consistent with scattering by gas or by very small (relative to visible wavelengths) particles or aggregates of small particles. The polarization signature is more interesting at northern high latitudes where high polarization is confined to the region inside the hexagon.

\begin{figure}%
\begin{center}
\figurebox{3.4in}{}{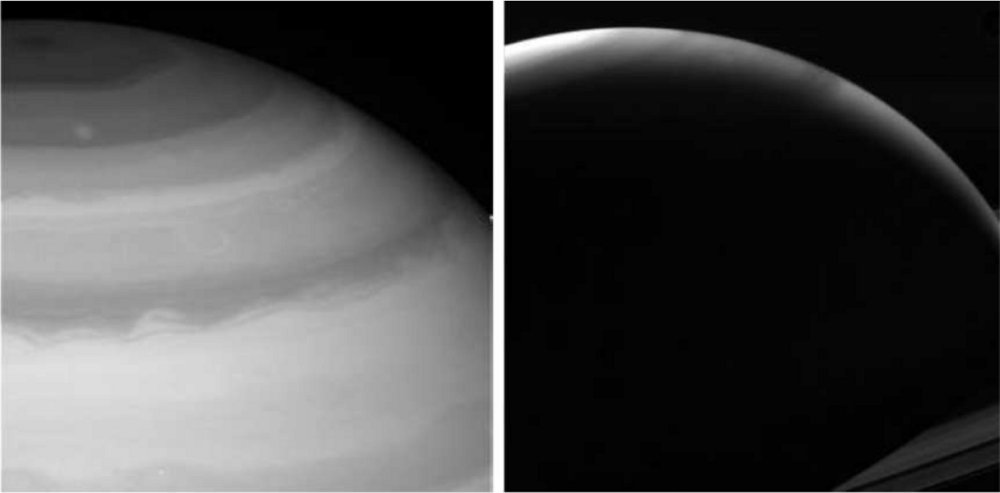}
\caption{These images in the Cassini ISS MT2 filter (727~nm methane band) obtained at low (2.6\degree) and high (151.3\degree) phase angles show the forward-scattering property of polar stratospheric aerosols. Saturn's spin axis is vertical in the left image and tipped to the left by 16 degrees in the right image. In both images the north pole is near the upper left corner. Figure from \cite{West_2014AOGS}.}
\label{Fig_PhaseAngle}
\end{center}
\end{figure}

In Fig.~\ref{Fig_HexPolarization} the material inside the hexagon appears dark in the images captured using the polarizer oriented with principal axis in the scattering plane \citep{West_etal_2015PolarimetryBook}. The different appearance in the two polarizer orientations is a signature of Rayleigh-type linear polarization. Some of this may be due to a larger optical path through molecular hydrogen and helium in this region (due to the absence of haze particles). However, methane-filter images at high phase angle show enhanced scattering from an aerosol layer at high latitudes (Fig.~\ref{Fig_PhaseAngle}). The combination of high polarization and strong forward scattering is also seen in the polar regions of Jupiter \citep{West_Smith_1991Titan-Jup_Aerosols} and points to a population of fractal aggregate particles in the stratosphere (probably near the 10-mbar pressure level). The small monomers, or primary particles, that combine to form the aggregate are responsible for the high polarization, while the large effective size of the aggregate is responsible for strong forward scattering. The details concerning particle optical depths, monomer size and average number of monomers in the aggregate remain to be worked out. 

\begin{figure}%
\begin{center}
\figurebox{3.4in}{}{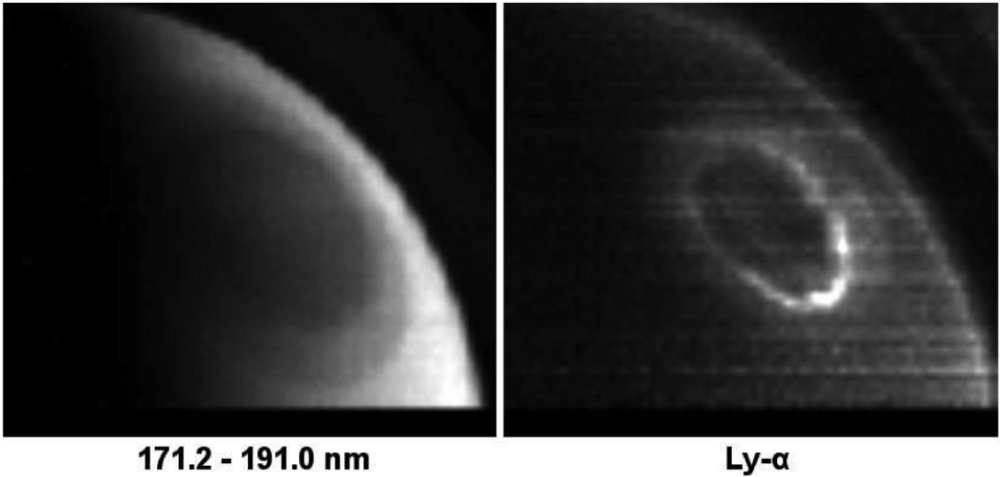}
\caption{UV imagery of Saturn's north polar region. The image on the left was constructed from a scan across the northern high latitude of Saturn by the Cassini UVIS intrument. It is the sum of spectral channels spanning the range 171.2-191.0~nm (reflected sunlight). On the right is the image in the hydrogen Ly-alpha line (121.6 nm) showing the auroral oval near 75\degree N. latitude (near the latitude of the hexagon). The images were from the same spectral cubes and are shown at the same scale. The UVIS slit moved across the horizontal direction; the streaks in the Ly-alpha image are from uncorrected residual flat field errors. Figure from \cite{West_2014AOGS}.}
\label{Fig:UVIShex}
\end{center}
\end{figure}

The latitudinal distribution of UV-absorbing haze provides another clue to the processes that control aerosol formation and distribution. UV-absorbing haze in the north is concentrated within the hexagon, as is the region of high polarization shown in Fig.~\ref{Fig_HexPolarization}. The region inside the hexagon is one of strong near-UV haze absorption as Fig.~\ref{Fig:UVIShex} shows \citep{West_2014AOGS}.

Acetylene gas also absorbs in the 171-191~nm region but the acetylene signature in the UV is nearly uniform over the image in Fig.~\ref{Fig:UVIShex} and does not contribute to the contrasts \citep{West_2014AOGS}. As discussed in Section~\ref{section:strat}, CIRS data reveal that acetylene concentration is increasing in the northern high latitudes in the infrared (Fig.~\ref{Fig:polarcxhy}); the UV signature is yet to be analyzed in conjunction with the CIRS data.

The confinement of the high polarization and UV absorption to the region inside the hexagon suggests that auroral processes lead to hydrocarbon production which ultimately leads to aerosol production in the high stratosphere in the form of fractal aggregate particles. Furthermore, the jet that forms the hexagon appears to provide a confining mechanism whereby particles produced within the hexagon are trapped. This mechanism is familiar in the terrestrial winter polar stratosphere where ozone is trapped within the winter polar vortex. Much of the Saturn data collected by the UVIS instrument remains to be analyzed, which could return results on stratospheric haze particle properties, photochemical processes that produce the particles, and dynamic transport processes that re-distribute the particles.

\section{Polar atmospheric dynamics}
\label{section:dynamics}

In this section, we review recent theoretical and modeling developments relevant to understanding giant planet polar atmospheric dynamics. As points of comparison, we use Jupiter and Saturn as the prototypical examples of the jovian-class planets (as opposed to the ice-giant class) that have been studied in detail. For Saturn specifically, we discuss the progress toward understanding the formation and maintenance of the hexagon and the polar vortices.

\subsection{Statement of the problem: explaining the three dynamical regimes on Jupiter and Saturn}
Spacecraft observations of Jupiter and Saturn since the early 1970s have identified three distinct dynamical regimes in the cloud-top winds. In the equatorial region, a fast, broad jetstream blows eastward where no vortices are found. In the mid-latitudes of both Jupiter and Saturn, many vortices exist between the numerous jetstreams that alternate in wind direction between eastward and westward --- Jupiter and Saturn have $\sim$30 and $\sim$15 jetstreams, respectively (Here, we define each of the eastward and westward peaks in the zonal wind profile as a \emph{jet}; nevertheless, such a definition inevitably carries a level of arbitrariness as exemplified by the opening sentence of \cite{Rhines_1994}: ``It is not easy to give a precise definition of a jet of fluid motion, but most of us know one when we see it.''). 

Closer to the poles, vortices become increasingly prevalent with latitude; however, Jupiter and Saturn critically differ in their atmospheric dynamics around the poles. On Jupiter, both the visible appearances and the dynamical characteristics change dramatically poleward of $\sim$60\degree~latitude compared to the lower latitudes. While the low- to mid-latitudes are characterized by alternating dark and bright bands whose boundaries loosely correspond to the peaks of the zonal jets, the high latitude regions lack clear banding and are instead marked by countless small vortices. Cassini observations of the northern high latitudes of Jupiter reveal a wind flow dominated by numerous vortices ranging in size from the limit of image resolution to thousands of kilometers \citep{Porco_etal_2003}. The motions of those high-latitude vortices are zonally organized as can be seen in the NASA public release movie PIA03452, and cloud tracking wind measurements such as that by \cite{Porco_etal_2003} is sensitive to vortex drifts. However, like the Great Red Spot and the White Ovals, those drifting vortices do not move at the background prevailing wind speed. We identify this vortex-filled dynamical regime as \emph{polar turbulence}. Saturn, on the other hand, maintains zonally organized cloud bands up to the poles and lacks Jupiter-like polar turbulence. As discussed in the previous sections, zonal structure of Saturn culminates in each of the hemispheres with a hurricane-like cyclonic vortex residing precisely at the poles. As of this writing, Jupiter's poles are yet to be observed with sufficient resolution to answer whether the planet harbors polar vortices --- our knowledge of Jupiter's poles is expected to be significantly updated by the upcoming Juno mission, which will go into orbit around Jupiter in 2016.

A comprehensive gas-giant planetary atmospheric dynamics model must explain these three dynamical wind regimes; (1) the vortex-free equatorial region with a fast, broad eastward jet, (2) the middle latitudes where stable vortices are embedded in the numerous robust zonal jets that violate the barotropic stability criterion, and (3) the vortex-dominated flow in the high latitudes. Such a model should also explain the difference between Jupiter and Saturn (and, eventually, the difference between the jovian and ice-giant classes of planets). In this review, we focus on our current understanding of the mechanism that separates the mixed jet-vortex flow regime to the polar turbulence, and the processes that differentiates the polar atmospheric dynamics of Jupiter and Saturn. 

\subsection{Transition from mid-latitude jets to polar turbulence}
A remarkable common property of the mid-latitude jestreams of Jupiter and Saturn is that they are extremely steady even though they violate the barotropic stability criterion at some latitudes. The spatial steadiness is signified by the observation that the longitudinal wind speed variation at a given latitude is much smaller than the variations of the zonal mean zonal wind in latitude \citep{Limaye_1986, Sanchez-Lavega_etal_2000}. Comparing the zonal mean wind profiles from the Voyager measurements and the Cassini measurements illustrates the spatial steadiness of the zonal jets in time, in which only minor changes in the locations and the speeds of those jets are found for both Jupiter (between 1979 and 2000; \citealt{Porco_etal_2003}) and Saturn (between 1980-81 and Cassini measurements since 2004; \citealt{Porco_etal_2005, Garcia-Melendo_etal_2011saturn}).

These stable mid-latitude jets are shaped by the effects of fast planetary rotation on the atmospheric turbulence. In giant planet atmospheres, the sources of turbulent motions such as thunderstorms and baroclinic instabilities can contain spatial scales smaller than the size of jets and vortices. In quasi-two-dimensional large-scale flows in stratified atmospheres, these small seeds of turbulence self-organize to form larger structures. The tendencies for the kinetic energy contained in small spatial features to be transferred to larger structures in two-dimensional fluid flows is often called the upscale (or inverse) cascade (early theoretical works include \citealt{Kraichnan_1967_InertialRanges, Kraichnan_1971_InertialRanges}, \citealt{Charney_1971}, and \citealt{Salmon_1980_BCinst-Gestroph}; also see textbooks such as \citealt{Salmon_1998} and \citealt{Vallis_2006}). When this process occurs on a fast-rotating planet such as Jupiter and Saturn, the resulting structures become elongated in east-west directions until they form zonal jets. This topic, the generation of zonal flow through geostrophic turbulence, has been extensively reviewed by \cite{Rhines_1979}, \cite{Rhines_1994}, \cite{Vasavada_Showman_2005} and \cite{Galperin_etal_2006}; in this review, we focus on the transition in the outcome of the upscale energy transfer from jet-dominated regime to a vortex-dominated flow.

The effects of planetary rotation on the upscale energy transfer was first explained in numerical experiments by \cite{Rhines_1975}. Using a barotropic model, Rhines showed that the upscale re-distribution of the kinetic energy in the spectral space results in structures elongated in the zonal directions with the north-south characteristic width
\begin{equation} \label{e:Rhines_Length}
L_\beta \approx (U/\beta)^{1/2}.
\end{equation}
where $\beta = df/dy > 0$, $y$ is northward distance, $f=2\Omega \sin(\phi)$, $\Omega$ is planetary rotation rate, and $\phi$ is latitude. $L_\beta$ is called the Rhines length (Note that the original form by Rhines is $L_\beta =\sqrt{2U/\beta}$, while the form presented in (Eq.~\ref{e:Rhines_Length}) has become more common in recent years). Recent studies further show that this upscale energy transfer is mediated by waves that transfer energy non-locally in the spectral space, i.e., the energy does not cascade between adjacent spectral modes \citep{Sukoriansky_etal_2007, Srinivasan_Young_2012}. Two-dimensional and quasi-two-dimensional geophysical turbulence remain an active area of investigation through numerical experiments (e.g., \citealt{Yoden_Yamada_1993, Chekhlov_etal_1996, Nozawa_Yoden_1997a, Huang_Robinson_1998, Yoden_etal_1999, Huang_Galperin_Sukoriansky_2001, Danilov_Gurarie_2004, Danilov_Gryanik_2004, Sukoriansky_etal_2007, Sayanagi_Showman_Dowling_2008}), laboratory experiments (e.g., \citealt{Read_etal_2007, Espa_etal_2010_JetExperiment, Slavin_Afanasyev_2012_BetaJetExperiment, DiNitto_Espa_Cenedese_2013ZonalJetExp, Smith_Speer_Griffiths_2014JetInLabAnnulus, Zhang_Afanasyev_2014BetaTurbExp, Read_etal_2015_LabJets}, and observational analyses (e.g., \citealt{Galperin_etal_2001PowerLaw, Galperin_etal_2004, Barrado-Izagirre_etal_2009SpatialSpec, Choi_Showman_2011JupTurbSpec, Galperin_etal_2014JupMacroTurb}). 

If the turbulent kinetic energy redistribution is indeed responsible for the generation of the mid-latitude zonal jets on Jupiter and Saturn, the transition from the mid-latitude flow regime to the polar turbulence must involve suppression of the effect. Using quasi-geostrophic (QG) models, \cite{Okuno_Masuda_2003} and \cite{Smith_2004} showed that the kinetic energy transfer can be suppressed when the Rossby deformation radius is small, as is the case in high-latitude regions of fast-rotating planets. For a single-layer fluid flow (e.g., shallow-water model), the Rossby deformation radius is $L_\mathrm{D}=\sqrt{gh}/f$, where $g$ and $h$ are the surface gravity acceleration and depth of flow, respectively. The barotropic model used in the original derivation by \cite{Rhines_1975} has $L_\mathrm{D}\thickapprox\infty$. When the deformation radius is finite, \cite{Okuno_Masuda_2003} showed that the Rhines length takes the form
\begin{equation} \label{e:Rhines_Length_with_LD}
  L_\beta' \approx \left(\frac{1}{L_\beta^{2}} - \frac{1}{L_\mathrm{D}^{2}}\right)^{-1/2}.
\end{equation}
When $L_\mathrm{D} \ll L_\beta$, (Eq. \ref{e:Rhines_Length_with_LD}) indicates that $L_\beta'$ becomes imaginary, and the kinetic energy transfer is suppressed \citep{Okuno_Masuda_2003, Smith_2004}. \cite{Theiss_2004} proposed that the boundary between the mid-latitude jet regime and the polar-turbulence region is marked by $L_\mathrm{D} \approx L_\beta$. \cite{Theiss_2006} further investigated the effect of the background flow on the value of effective $\beta$, essentially taking account of the change in the Rossby wave frequency due to the zonal flow. \cite{Theiss_2006} and \cite{Penny_etal_2010SatVorts} showed that the distributions of vortices on Jupiter and Saturn are consistent with the suppression of the upscale kinetic energy transfer by the background zonal flow. The difference in the high-latitude atmospheric dynamics between vortex-filled Jupiter and zonally organized Saturn could be explained by the relative scales between the deformation radius $L_\mathrm{D}$ and the Rhines length $L_\beta'$ on each of the planets --- $L_\mathrm{D}$ is indeed generally smaller on Jupiter than on Saturn in low- and mid-latitudes \citep{Read_etal_2006, Read_etal_2009}; however, $L_\mathrm{D}$ and $L_\beta'$ remain to be estimated for high-latitude regions of either planet.

Studies using shallow-water (SW) models show that the suppression of the upscale kinetic energy transfer under small $L_\textrm{D}$ also occurs when the ageostrophic effects are fully included. Simulations of freely-evolving \citep{Cho_Polvani_1996PhysFl, Iacono_etal_1999PhysFl, Kitamura_Ishioka_2007} and forced \citep{Showman_2007, Scott_Polvani_2007} turbulence showed that the flows in such systems form multiple zonal jets in low latitudes accompanied by vortex-dominated regions in the high latitudes. Note that, under horizontally uniform stratification, $L_\textrm{D}$ decreases with latitude as $f$ increases. In these 1-layer simulations, a clear critical latitude divides the jets and the vortex-dominated regions. However, these shallow-water models do not explain two aspects of the zonal jets on Jupiter and Saturn. First, under small $L_\textrm{D}$ like those of Jupiter and Saturn, their equatorial jet becomes westward \citep{Showman_2007, Scott_Polvani_2007, Kitamura_Ishioka_2007}, which is the opposite of Jupiter or Saturn. Second, they do not reproduce the mid-latitude flow regime in which small vortices and stable zonal jets coexist. 

A mixed jet-vortex flow regime like that of Jovian mid-latitudes can be reproduced using three-dimensional multi-layer numerical models that solve the primitive equations. \cite{Sayanagi_Showman_Dowling_2008} demonstrated that, in freely-evolving simulations of geostrophic turbulence on a beta-plane, the outcome of the turbulent self-organization transitioned from the mixed jet-vortex regime to the vortex-dominated regime as the deformation radius $L_\textrm{D}$ was decreased. Their experiment can be seen as a three-dimensional counterpart to the SW simulations by \cite{Cho_Polvani_1996PhysFl} and \cite{Iacono_etal_1999PhysFl}. To date, studies of the effects of forced three-dimensional turbulence on giant planet atmospheres mostly focus on jet formation through baroclinic instabilities, and few studies focus on high latitudes. Nevertheless, a gradual increase in the vortical activities toward the poles can be discerned in the results by \cite{Williams_2003a} and \cite{Liu_Schneider_2010JovJets}. Unlike the 1-layer SW models, these many-layer three-dimensional models can have multiple baroclinic deformation radii that become relevant in their dynamics (e.g., \citealt{Achterberg_Ingersoll_1989}). \cite{Sayanagi_Showman_Dowling_2008} speculated that a mixed jet-vortex regime occurs when the Rhines length is smaller than some, but not all, deformation radii; in such a scenario, the upscale energy transfer would produce vortices at those deformation radii smaller than the Rhines length and co-exist with jets. Much work remains to be done before we gain a full understanding of the effect of three dimensionality on the jet formation and the upscale energy transfer.

It is worth noting that some recent studies show that the upscale energy transfer may not be solely responsible for the generation of the zonal jets (for example, \citealt{Sukoriansky_etal_2007, Scott_Tissier_2012_LargeMixJets, Verhoeven_Stellmach_2014CompressBetaJet}). However, we limit the scope of our current discussion to the transition of flow regime from the mixed jet-vortex regime to the polar turbulence through the suppression of the upscale energy transfer.

\subsection{Models of the hexagon}

\begin{figure*}%
\begin{center}
\figurebox{6.85in}{}{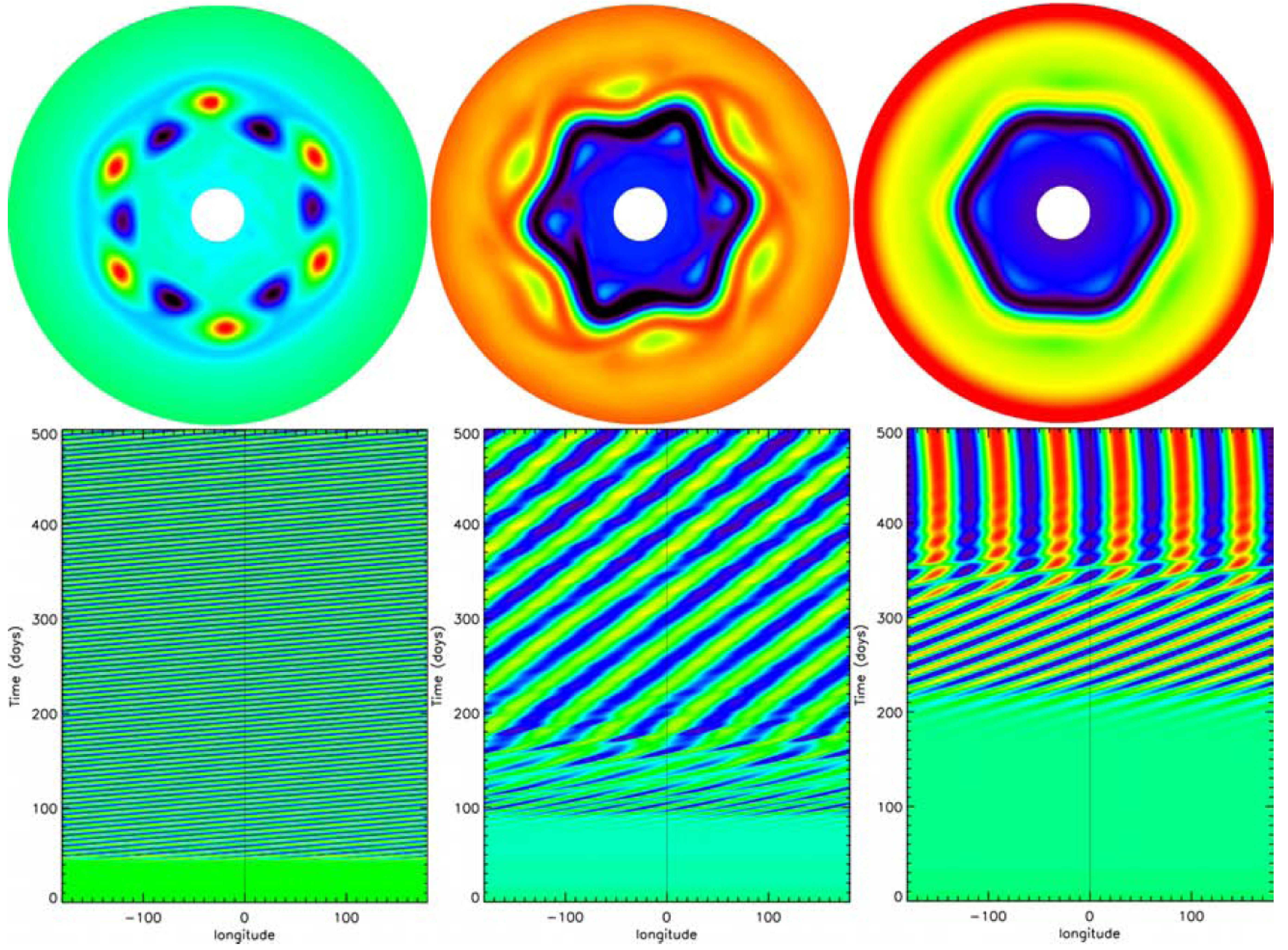}
\caption{\cite{Morales-Juberias_etal_2015Hexagon} showed that, depending on the vertical wind structure, a barotropically unstable jetstream around the pole can relax into a vortex street (left), a meandering jet with a star-shaped path (middle) and a meandering jet with a polygonal path (right). The top panels show maps of potential vorticity at the 1-bar level, in which the value increases in order of red-green-blue. The bottom panels are Hovm\"{o}ller diagrams that depict the propagation of the hexagonal disturbance in time, which shows that the polygonal flow pattern is nearly stationary.}
\label{Fig:HexModel2015}
\end{center}
\end{figure*}

As discussed in Section~\ref{section:hex}, the hexagon was first discovered by \cite{Godfrey_1988} in images captured by Voyager in 1980-81. The key characteristics of the feature that must be explained are; (1) its hexagonal outline follows the path of a fast eastward jetstream that blows at around 75\degree N PC latitude \citep{Godfrey_1988, Baines_etal_2009poles, Sanchez-Lavega_etal_2014_Hexagon, Antunano_etal_2015_SatPoleDyn}; (2) the feature does not accompany large vortices (evident in the measurements by \citealt{Antunano_etal_2015_SatPoleDyn}); (3) the hexagonal pattern's longitudinal drift is extremely slow in the System~III reference frame (0.8$\pm$1.1 m~s$^{-1}$ as measured by \citealt{Godfrey_1988} using Voyager data and $-$0.036$\pm$0.004~m~s$^{-1}$ by \citealt{Sanchez-Lavega_etal_2014_Hexagon} using combination of Cassini and ground-based data); and (4) the potential vorticity gradient across the hexagon jet is substantially steeper than the surrounding region \citep{Read_etal_2009}, i.e., it forms a \emph{PV front}.

A substantial body of literature about meandering geophysical jets accompanied by PV fronts can be found in studies of Earth's meandering ocean currents (e.g., \citealt{Allen_Walstad_Newberger_1991, Flierl_1999}) and mid-latitude atmospheric jets (e.g., \citealt{Polavarapu_Peltier_1993b, Swanson_Pierrehumbert_1994}). A jetstream with a sharp PV front can develop finite amplitude instability (i.e., an instability which amplitude growth is damped by nonlinear effects); \cite{Hart_1979AnRev} is an early review on the topic that introduced the idea. To the best of our knowledge, \citet{Flierl_Malanotte-Rizzoli_Zabusky_1987} conducted the first numerical experiments to demonstrate that an unstable jet can develop instabilities, which subsequent amplitude growth halts nonlinearly in some cases and finds a stable configuration. Using a barotropic model, they tested the stability of various initial flows that contain single-mode perturbations and examined the instability growth. Their results demonstrate that, depending on the initial flow, the flow can equilibrate in either (1) a vortex street; (2) a stably meandering jet with a monochromatic wavelength; or (3) a jet with chaotic unsteady meandering. Here, a vortex street is a fluid flow structure that has a chain of vortices with alternating signs of vorticity (e.g. \citealt{Karman_1921}). Many numerical models and laboratory experiments have examined each of these outcomes as summarized by \cite{Sayanagi_Morales-Juberias_Ingersoll_2010} in the context of Saturn's ribbon jet at 41\degree N PC (47\degree N PG) latitude, which exhibits chaotic meandering \citep{Sromovsky_etal_1983, Godfrey_Moore_1986}. The hexagon jet is a case of a steadily meandering zonal jet with a monochromatic wavelength; the geometry of the the meandering appears polygonal when viewed from above due to its high latitude.

\adjustfigure{85pt}

Numerical and laboratory experiments to date have demonstrated a vortex street and a meandering jet can both have polygonal flow path even though their vorticity structures are different. In a vortex street, each of the vortices has a local peak of relative vorticity surrounded by closed streamlines. In comparison, a meandering jet does not accompany closed streamline vortices and the streamlines are instead roughly parallel to the jet's path. The emergence of a polygonal structure associated with a meandering, barotropically unstable zonal jet was first shown in a laboratory experiment by \cite{Sommeria_etal_1989LabMeander}, in which they produced a pentagonal flow in a rotating annular tank that accompanies a sharp potential vorticity front. \cite{Solomon_etal_1993LabJetInst} and \cite{Marcus_Lee_1998} performed laboratory and numerical experiments to demonstrate that shear instability can equilibrate to form a vortex street. In particular, \cite{Marcus_Lee_1998} showed that, when the vortex street consists of 6 pairs of interlocking cyclones and anticyclones around the jetstream, the path of the jet resembled the hexagon; however, unlike these vortex street flows, the Saturnian jet at 75\degree N PC latitude lacks strong vortices adjacent to the jet.

\cite{BarbosaAguiar_etal_2010} applied these previous results to parameter regimes relevant to the hexagon through linear stability analysis and laboratory experiments. They studied a flow with parameters relevant to the 75\degree N hexagon jet ($L_\mathrm{D} = N_\mathrm{B}H/f_0$, where $N_\mathrm{B}$ is the Brunt-V\"{a}is\"{a}l\"{a} frequency, and $H$ is the atmospheric scale height) and showed that, when Rossby deformation radius $L_\mathrm{D} = \infty$, the maximum growth rate occurs for a zonal wavenumber $k_\textrm{max} = 13$. However, when $L_\mathrm{D}$ = 2500~km, the most unstable modes become $k_\mathrm{max} \sim$ 5-6 and the instability growth rate also decreases, meaning that the flow is more stable. For $L_\mathrm{D} \leqq$ 2000~km, the large-scale instability and the growth rate increases monotonically with the wavenumber. \cite{BarbosaAguiar_etal_2010} applied the result of this analysis to conduct fluid dynamic laboratory experiments in a rotating tank (similar setup to those by \cite{Sommeria_etal_1989LabMeander}, \cite{Solomon_etal_1993LabJetInst} and \cite{Marcus_Lee_1998} but with a different forcing mechanism) to simulate the hexagon formation mechanism to show that the jet equilibrates into a wavenumber-6 mode, and demonstrated that barotropic instabilities lead to a meandering jet that forms a hexagon. One aspect of \cite{BarbosaAguiar_etal_2010}'s experimental result that is not consistent with the observation is the presence of large vortices that sandwich the meandering jet --- such vortices are absent in the observed wind profile \citep{Antunano_etal_2015_SatPoleDyn}. \cite{Morales-Juberias_etal_2011, Morales-Juberias_etal_2015Hexagon} discuss that \cite{BarbosaAguiar_etal_2010}'s results resemble a vortex street.

\cite{Morales-Juberias_etal_2011, Morales-Juberias_etal_2015Hexagon} studied the instabilities arising in an unstable jet encircling the north pole of Saturn using nonlinear three-dimensional atmospheric dynamics models. Their first study showed that, when the jet is deep-seated (i.e., when assumed to have a negligible vertical shear below the observed cloud-top), the jet relaxed into a vortex street. They demonstrated that the dominant wavenumber of the final relaxed state depends on the width of the initial jet by tuning the wavenumber between 3 and 8. However, all their outcomes were vortex streets, and the feature propagated in longitude much faster than the observed behavior of the Saturnian hexagon. Their second study demonstrated that, when the initial jets were shallow (i.e., their wind speed decreased with depth below the observed cloud top such that it approached zero near the cloud condensation level at 10~bar), the instabilities relaxed the jets into a steady, near-stationary meandering state accompanied by a PV front (Fig.~\ref{Fig:HexModel2015}), thus reproducing all of the four key characteristics observed in the hexagon noted earlier. \cite{Morales-Juberias_etal_2011} also showed that a vortex street model cannot explain the slow propagation rate of the observed hexagon. The strong vertical shear of the jet presumably makes the instabilities in \cite{Morales-Juberias_etal_2015Hexagon}'s model more baroclinic rather than barotropic, which would make the dynamics more like Saturn's ribbon wave \citep{Godfrey_Moore_1986, Sayanagi_Morales-Juberias_Ingersoll_2010}; however, \cite{Morales-Juberias_etal_2015Hexagon} do not analyze the energetics of the flow and the details remain unclear.

\cite{Morales-Juberias_etal_2015Hexagon}'s results suggest that Saturn's 75\degree N jet must be a shallow atmospheric structure. \cite{Liu_Schneider_2010JovJets}'s numerical simulations, which also assumed shallow jet structures, also produced meandering polygonal high-latitude jets for Saturn. Such a zonal jet with high vertical wind shear must accompany a horizontal temperature gradient through the thermal wind balance, but temperature measurements from these relevant depths below the cloud tops do not yet exist --- the results by \cite{Morales-Juberias_etal_2015Hexagon} are within the uncertainties of the observed temperature structure by \cite{Fletcher_etal_2008_CIRS_PoleHex}. Measurements of Saturn's high-order gravitational moments \citep{Liu_Schneider_Fletcher_2014SatJetDepth} may be able to test this hypothesis that the 75\degree N jet is shallow. This shallow-jet scenario is not consistent with a deep-rooted jet advocated by \cite{Sanchez-Lavega_etal_2014_Hexagon}; this disagreement could be resolved by the gravity measurements, which are sensitive to depth of jets, during Cassini's Grand Finale orbits. \cite{Morales-Juberias_etal_2015Hexagon} also showed that an identical jet centered at 70\degree N does not develop meandering, suggesting that the slightly equatorward center latitude of the southern jet at 70\degree S PC is responsible for the lack of polygonal meandering in the south.

\begin{table*}
\begin{center}
\caption{Comparison of the Saturnian Polar Vortices (Saturn PVs) with other vortices in the solar system based on \cite{Dyudina_etal_2009} with updates. $^{(1)}$ This assumes that the eastward zonal wind peak at $\sim$72\degree S in the measurement by \cite{Sromovsky_etal_2001_PartIII_NepWind} is the polar-most zonal wind peak, even though the wind structure poleward remains unresolved. $^{(2)}$ \cite{Luszcz-Cook_etal_2010Nep} showed that Neptune's south polar vortex was split into two in July~2007.}
\begin{tabular} {l l l l l l l l l}
  \hline \hline
  Parameters & Saturn PVs & Hurricane & Venus PVs & Earth PV & GRS & Neptune Pole\\ 
  \hline
        Radius (km) & 2000 & 10-100 & $\sim$5000 & $\sim$2500 & 7000-20000 & $\sim$7500$^{(1)}$ \\
        Radius (\% planetary radius) & 3 & 0.15-1.5 & $\sim$85 & $\sim$40 & 10-30 & $\sim$30\\
        Lifetime & $>$11~years & weeks & years & seasonal & $>$150~years & unknown \\
        Max Wind Speed (m~s$^{-1}$) & 150 & 85 & 35 & 50 & 120-190 & 250$^{(1)}$ \\
        Temperature Anomaly & warm & warm & warm & cold & cold & warm \\
        Vorticity & cyclonic & cyclonic & cyclonic & cyclonic & anticyclonic & cyclonic \\
        Eyewall Clouds & yes & yes & unknown & no & unknown & unknown \\
        Fixed to Pole & yes & no & yes & yes & no & yes$^{(2)}$ \\
        Convective Clouds on periphery & yes & yes & no & no & yes & unknown \\
        Bounded by rigid surface & no & yes & no & yes & no & no \\ 
        $| \zeta /f | \sim R_{\mathrm{o}}$ & 1 & 100 & 100 & 1 & 0.2 & 0.25$^{(1)}$ \\
        \hline \hline
 \end{tabular}
 \label{Tab:VortCompare}
\end{center}
\end{table*}

\subsection{Models of polar vortices}
The polar vortices of Saturn show many similarities to other vortices elsewhere in the solar system. Table~\ref{Tab:VortCompare} compares the dynamical characteristics of the Saturnian polar vortices to a few other notable vortices. In the table, Saturnian polar vortices (Saturn PVs) are compared with terrestrial hurricanes \citep{Anthes_1982book, Emanuel_2003AnnRev}, Arctic and Antarctic polar vortices on Earth (Earth PV) \citep{Nash_etal_1996_Earth-Polar-Vortex}, Jupiter's Great Red Spot (GRS) \citep{Simon-Miller_etal_2002_GRS, Choi_etal_2007, Simon_etal_2014_GRS-Shrink}, south polar vortex on Venus (Venus PV) \citep{Piccioni_etal_2007VenSpole, Limaye_etal_2009VenusPolarVort, Patsaeva_etal_2015VenusWind}, and the warm spot at the south pole of Neptune \citep{Orton_etal_2012NepPole, Sromovsky_etal_2001_PartIII_NepWind}. The ``radius'' refers to the major axis of the area enclosed within the maximum winds (\citealt{Emanuel_2003AnnRev} defines the radius of a cyclone as the radius where the wind speed becomes indistinguishable from the background wind; however, we use the radius of the maximum wind here because, for polar vortices, vortex wind speeds and background zonal winds are impossible to distinguish). For the Saturnian PVs and terrestrial hurricanes, the radius of the eye is used. The temperature anomaly refers to the emission temperature at the cloud top relative to its surrounding, which may have different sign higher or lower in the atmosphere. The ``unknown'' eyewall clouds may be blocked from the view by higher clouds. The vorticity ratio in the table $\zeta/f$ is effectively the Rossby number $R_\textrm{o}$. 

The table reveals that the Saturnian polar vortices, terrestrial hurricanes, and Venusian polar vortices share the warm-cored, cyclonic properties. In Table~\ref{Tab:VortCompare}, Jupiter's Great Red Spot is classified as anticyclonic and cold-cored based on its observed cloud-top properties; however, its structure at depth remains unknown (note that the circulations in terrestrial hurricanes transition vertically from cyclonic near the surface to anticyclonic aloft; \citealt{Frank_1977_HurricaneStructI}). 

In contrast to the warm-cored Saturnian polar vortices, Earth's polar vortices are cold cored near the surface \citep{Holton_2004}. The cold surface temperature occurs because radiative effects are more dominant in the troposphere and stratosphere than horizontal transport and mixing of heat. Consequently, in winter, the pole cools radiatively, and a stronger vertical shear develops in the eastward winds of the circumpolar vortex to balance the horizontal temperature gradient via the thermal wind equation. However, higher up in the mesosphere, radiative effects are smaller and the adiabatic heating associated with subsidence over the winter pole becomes dominant, and temperatures become relatively warmer than those at lower latitudes. As a result, the polar vortex decays with altitude in the mesosphere. Indeed, the mesopause ($\sim$90 km) is warmest over the winter pole of Earth. (The situation on Titan is similar, except that it is the stratopause that is warm over the winter pole, and meridional contrasts actually decrease at higher altitudes).

Unlike Earth's polar vortices, Saturn's polar vortices probably do not have the vertical dipolar structure described in the previous paragraph. Saturn's internal heat source tends to reduce lateral variations in entropy at depth (e.g., \citealt{Allison_2000_WindyThermocline}), and radiative damping times in the lower troposphere are too large to form a cold core over the poles in winter. Given that warm cores are observed at both poles in the upper troposphere (150 mbar) apparently independent of season (Fig.~\ref{Fig:polarmaps}; \citealt{Fletcher_etal_2008_CIRS_PoleHex, Fletcher_etal_2015SaturnPoles}), it is clear that the terrestrial polar analog does not apply. An analogy to hurricane would offer a way around this problem, in which cyclonic vorticity develops via low-level convergence and uplift, with maximum ascent in the eyewall. In terrestrial hurricanes, the swirling winds are maximum near the surface and decrease with altitude, and the warm temperatures within the eye, implied by the thermal wind equation, are maintained by a weaker descending flow in the eye (e.g., \citealt{Frank_1977_HurricaneStructI, Frank_1977_HurricaneStructII}).

The morphological similarity between the Saturnian polar vortices and the terrestrial tropical cyclones (such as hurricanes) are especially notable. First, both examples have a clear hole in the clouds (``eye'') at the center that allows space-based remote-sensing instruments to probe deeper into the atmosphere (down to the surface for the case of Earth). Second, both types of vortices have cyclonic vorticity and swirling cloud patterns that reflect the wind shear surrounding the central eye. However, a problem in drawing an analogy between Saturnian polar vortices to terrestrial hurricanes is that the dominant forcing necessary for their generation is the latent heat release from the warm ocean surface underlying the vortex \citep{Anthes_1982book, Houze_1993book, Emanuel_2003AnnRev}, but there is no analogous process on Saturn.

A mechanism that has been proposed for the formation of the Saturnian polar vortices involves the tendency for cyclonic vortices to drift poleward called the $\beta$-drift. The $\beta$-drift is caused by the difference in the potential vorticity of a vortex with its surrounding; a vortex drifts toward a direction where the environmental potential vorticity matches more closely with that of the vortex (an analogy could be drawn with motion caused by buoyancy force, in which a parcel of fluid rises or sinks toward a direction where the environmental density matches more closely with that of the parcel). The $\beta$-drift is the main mechanism that drives terrestrial tropical cyclones poleward, and has been studied extensively (e.g., \citealt{Reznik_1992beta-drift, Nycander_1993beta-drift, Sutyrin_etal_1994beta-drift, Sutyrin_Morel_1997beta-drift, Montgomery_Moller_Nicklas_1999, SaiLapLam_Dritschel_2001_beta-drift}). Elsewhere in the solar system, the $\beta$-drift mechanism is responsible for the equatorward drift of the anticyclonic dark spots on Uranus \citep{Hammel_etal_2009UraDarkSpot, dePater_etal_2011Berg} and Neptune \citep{Lebeau_Dowling_1998, Stratman_etal_2001}.

The accumulation of cyclonic vorticity can be seen in the outcomes of freely-evolving and forced turbulence simulations of geostrophic turbulence \citep{Cho_Polvani_1996PhysFl, Scott_Polvani_2007, Scott_2010turb-book}, though the formation of polar vortices was not the focus of these studies. Taking hints from these studies, \cite{Scott_2011PolarCyclone} systematically tested the formation of polar vortices through the $\beta$-drift mechanism by running freely-evolving simulations of polar turbulence using a one-layer quasi-geostrophic shallow-water model. When the flow is initialized with a random distribution of small cyclones and anticyclones, the cyclones indeed $\beta$-drifted poleward as expected and merged at the pole to form an intense polar cyclone. \cite{ONeill_Emanuel_Flierl_2015NatGeo, ONeill_Emanuel_Flierl_2016JAS} further tested the idea with simulations of forced-turbulence using a two-layer shallow-water model (i.e., not quasi-geostrophic) to show that the mechanism works in generating a polar vortex when forcing, ageostrophic effects, and some three-dimensional effects are included. 

\cite{ONeill_Emanuel_Flierl_2015NatGeo, ONeill_Emanuel_Flierl_2016JAS} also showed that a strong polar cyclone emerges only when the deformation radius $L_\mathrm{D}$ is relatively large ($\sim$1/20 of the planetary radius) and cumulus convection is energetic enough. When $L_\mathrm{D}$ is reduced to $\sim$1/40 of the planetary radius (a Jupiter-like regime), a polar vortex does not form but instead weak jets emerge like those in the high-latitude regions of Jupiter. When $L_\mathrm{D}$ is held at $\sim$1/20 of the planetary radius and cumulus convective energy flux is reduced, a transient polar vortex emerged similar to that observed on Neptune by \cite{Luszcz-Cook_etal_2010Nep}. \cite{ONeill_Emanuel_Flierl_2015NatGeo, ONeill_Emanuel_Flierl_2016JAS}'s results suggest that differences in the polar atmospheric dynamics between Jupiter, Saturn and Neptune are controlled by the deformation radius and cumulus convective energy flux; based on this result, they predict that Jupiter's poles have weak jets and lack a Saturn-like polar vortex. This prediction of reduced zonal organization for Jupiter than for Saturn is consistent with the prediction that the region affected by polar turbulence (described by a region of $L_\mathrm{D} \ll L_\beta$ following Eq.~\ref{e:Rhines_Length_with_LD}) should expand when $L_\mathrm{D}$ is reduced. 

On Earth, tropical cyclones require a warm sea surface to form, but their intensification follows a process similar to that tested by \cite{Scott_2011PolarCyclone}. When a terrestrial tropical cyclone becomes intense enough, it creates a local peak in the potential vorticity field such that it generates a local \emph{effective} $\beta$-effect linked to the local gradient of potential vorticity. The process involving local effective-$\beta$ is analogous to that of the planetary-$\beta$, which is the consequence of the gradient in the planetary vorticity, and the results are similar; when cumulus convection generates eddies inside of a terrestrial tropical cyclone, it causes an inward flux of cyclonic vorticity, and intensifies the cyclone \citep{Montgomery_Enagonio_1998}. Another implication of the locally modified effective $\beta$ is the propagation of ``vortex Rossby waves.'' As a cyclonic vortex has a potential vorticity local maximum at its center, Rossby waves propagate azimuthally around the center of vortex and transport the azimuthal momentum component. \cite{Montgomery_Kallenbach_1997} proposed the link between the excitation and absorption of the vortex Rossby waves to the intensification of hurricanes, which was further studied by \cite{Montgomery_Moller_Nicklas_1999} and \cite{Moller_Montgomery_2000}. The intensification of terrestrial hurricanes and subsequent formation of an eye has also been linked to the vortex Rossby waves \citep{Wang_2002HurricaneIntensification}. Also, similar to the pumping of zonal jets on Jupiter (e.g., \citealt{Salyk_Ingersoll_etal_2006}) and Saturn \citep{DelGenio_etal_2007}, eddy momentum flux divergence is proposed as an important diagnostic quantity in tropical cyclone intensification \citep{Nolan_Ferrall_1999HurricaneEddyForcing}. The vortex Rossby waves and eddy momentum flux divergence should also play a role in the Saturnian polar vortices' intensification and eye formation, but these effects are yet to be studied for Saturn.

\section{Discussion and unanswered questions}
\label{section:conclusions}

Over the past decade, the Cassini Orbiter has discovered a wealth of information about Saturn's poles, including the following.\\
1. The existence of large cyclonic vortices centered over both poles, covering $\sim$10\degree (9,500~km) and $\sim$15\degree (14,000~km) of latitude in radius for the north and south polar vortices, respectively. (Here, a vortex radius is defined by the region poleward of the polar-most minimum in the eastward wind.) \\
2. These polar cyclonic vortices extend over $\geqq$100~km in depth, as revealed in (i) the classic latitudinal vortex structure observed in the tropospheric zonal winds down to at least the 2-bar level and (ii) in the latitudinal thermal temperature structure throughout the upper stratosphere up to at least the 1-mbar level. \\
3. The polar vortices have morphological and dynamical similarities to terrestrial hurricanes. \\
4. Tropospheric winds in both polar vortices exceed 120~m~s$^{-1}$ within 1~degree of the pole, decreasing steeply to calm winds at the poles. Maximum winds of 136~m~s$^{-1}$ and 174~m~s$^{-1}$ occur near 88.3\degree N PC (88.6\degree PG) and 87.5\degree S PC (88.0\degree S PG), respectively. \\
5. While seasonal changes in stratospheric temperatures are apparent in polar regions, the classical latitudinal thermal structure of cyclonic polar vortices has persisted over each pole for 1/3 Saturn year since their discovery by Cassini. \\
6. High-speed jets border the polar vortices in both hemispheres: The Hexagon centered at 75\degree N PC (78\degree N PG) with a maximum wind speed of $\sim$130~m~s$^{-1}$, and a southern jet at 70\degree S PC (73\degree PG) with a maximum wind speed of $\sim$87~m~s$^{-1}$.\\
7. The north polar hexagon has now been consistently observed for over 35 years, from its discovery in Voyager images acquired in 1980 through Cassini and high-resolution ground-based images acquired today, in the Spring of 2015, including many acquired with newly-developed techniques by amateur astronomers (\emph{c.f.} Chapter 14).\\
8. The hexagon is a meandering jet that is not accompanied by large accompanying vortices.

Despite these revelations, fundamental questions remain, including the following.\\
A. Why does Saturn display intense polar vortices, complete with ``hurricane eyes'' on both poles? Does Jupiter, its fellow gas giant, lack such vortices as predicted by \cite{ONeill_Emanuel_Flierl_2015NatGeo, ONeill_Emanuel_Flierl_2016JAS}? \\
B. What sustains the hexagonal and the other high-speed jets? \\
C. How deeply do the hexagon and the polar vortices extend below the visible cloud-top? \\
D. How do seasonal effects affect the cloud morphologies and meridional circulation associated with the polar vortices and the hexagon? \\
E. What are the depth of cloud layers, each exhibiting different morphology, sensed at different wavelengths? \\
F. What is the role of hydrocarbon photochemical products, including aerosols, on the radiative energy balance? \\
G. What are the properties (e.g., altitude, density, radius, composition) of the haze aerosol particles, which may be fractal aggregates of many particles? \\
H. How does the ion chemistry associated with aurorae work, and their effects on the haze particle production?

\subsection{Outlook: end of Cassini mission}
To help answer such questions, the Cassini Orbiter will provide some of its most detailed and comprehensive imagery and spectra yet during and just prior to the Grand Finale Phase in late 2016 and 2017. In November 2016, while flying a series of inclined orbits that begin in February 2016, Cassini encounters Titan, reducing its closest approach distance from $\sim$279,000~km above the clouds of Saturn to $\sim$158,000~km. This provides a 24~hour period for clear, sunlit viewing of the north polar region. During the orbiter's inbound trajectory, its sub-spacecraft latitude will first increase from 45\degree N at an altitude of $\sim$809,000~km to a peak latitude of $\sim$60\degree N (PC); the periapsis altitude will be $\sim$290,000~km above 55\degree N latitude, from where the north polar region will be severely foreshortened. Encounters on November~18 and 27, 2016 from similar viewing geometries allow detailed investigations of polar winds and structure by the ISS Camera, the VIMS visual-near-infrared spectral mapper, and the CIRS thermal spectral mapper. During the orbiter's outbound trajectory after each periapsis pass, over a 10-hour period, the spacecraft will maintain favorable observing geometries of the southern high latitudes when its sub-spacecraft latitude first varies from $\sim$55\degree S to $\sim$60\degree S, and then equatorward to $\sim$45\degree S from altitudes of $\sim$205,000 to $\sim$450,000~km, enabling VIMS and CIRS instruments to capture thermal maps of the south polar region under nighttime conditions. 

On November 29, 2016, another Titan encounter lowers the closest approach altitude distance to $\sim$90,000 km, setting up the first phase of the Grand Finale mission. For twenty orbits, the spacecraft will maintain highly eccentric orbits that pass over the north polar region; as the spacecraft approaches the periapsis, its altitude decreases from $\sim$600,000~km to $\sim$165,000~km over a $\sim$15.2-hour period as the sub-spacecraft latitude varies from 55\degree N to 45\degree N. The south polar region is then viewed from distances ranging from $\sim$126,000 to $\sim$300,000~km over $\sim$7.4~hours. Presently, Cassini is scheduled to observe Saturn during four of theses twenty orbits with periapses on January 16, February 14, March 7, and March 29, 2017. 

A final Titan pass on April 22, 2017 starts the second and final stage of the Grand Finale mission phase, sending the Cassini orbiter to a periapsis altitude of less than 3000~km over the Saturnian clouds. For the next 22~encounters, until the final plunge of the spacecraft into Saturn on September 15, 2017, the Cassini Orbiter obtains the highest resolution images of Saturn ever obtained. The orbiter will observe the north pole while its sub-spacecraft latitude varies from 55\degree N to 45\degree N from altitudes of $\sim$270,000 to $\sim$42,000~km over $\sim$4.5~hours. The southern high latitudes will be observed while the sub-spacecraft latitude varies from 55\degree S to 45\degree S from an altitude of $\sim$26,000 to $\sim$160,000~km over $\sim$2.3~hours. Cassini is currently scheduled to observe the poles extensively during four of the twenty orbits, on April 26, June 29, August 1 and August 20, 2017. From within 50,000~km, the VIMS instrument will produce spectral maps of the hexagon and north polar regions with individual pixel scale of less than 25~km at the cloud-top, CIRS will obtain thermal maps with better than 13-km sampling, and ISS will produce visual images with just 300-meter pixel resolution. However, due to the swift movement of the spacecraft (exceeding 34 km~s$^{-1}$, or 122,000~km~hour$^{-1}$), only a few such high-resolution images are expected to be captured during each pass. Nevertheless, virtually all imagery taken within the 270,000 km start of each polar pass will be of higher spatial resolution than heretofore obtained during the mission. 

The final phase of the Cassini mission will address many of the unanswered questions by extending the temporal observational coverage that stretches between almost two solstices, and by obtaining unprecedented high spatial resolution data enabled by the low-altitude orbits during the final phase of the mission. We expect that the extended seasonal coverage will address those questions related to seasonal forcing on the thermal structure and aerosol distribution, and confirm the predicted formation of the warm north polar stratospheric vortex, mirroring that observed over the south pole at the start of the mission. The questions on the vertical structure of the zonal jets may be answered through the analysis of the gravity field during the Grand Finale orbits. Higher resolution images to be returned during the final phase of the mission will also help reveal the detailed structure of the hexagon and the polar vortices. Other questions on the details of light-scattering properties of aerosol particles, auroral ion chemistry, and radiative energy balance will require further theoretical and laboratory analyses. Cassini will perform the Grand Finale polar observations while the Juno mission will make similar observations of Jupiter; together, these missions are expected to significantly update our knowledge of polar atmospheric science of Jupiter and Saturn.

In addition, many images captured by VIMS, UVIS and ISS have not been analyzed, which could be used to reveal, for example, quantitative information on the haze and clouds particles altitude, density, radius, composition --- continued analysis of the enormous data collected by the Cassini mission must be part of the effort to address many of these unanswered questions for many years to come. The Cassini mission's decade-long tour of the Saturnian system has rewritten our understanding of various processes that shape Saturn's polar atmosphere. The remaining years of the mission promise to return further data that will add to our understanding of the ringed planet.

\section*{Acknowledgements} 
Part of this work was performed by the Jet Propulsion Laboratory and was supported by the Cassini project. Sayanagi was supported in part by grants from NASA (Outer Planet Research Program NNX12AR38G, Planetary Atmospheres Program NNX14AK07G, and Cassini Data Analysis Program NNX15AD33G) and NSF (Astronomy \& Astrophysics Program 1212216). Fletcher was supported by a Royal Society Research Fellowship at the University of Leicester. S{\'a}nchez-Lavega was supported by the Spanish MICIIN AYA2012-36666 with FEDER funds, Grupos Gobierno Vasco IT765-013, and UPV/EHU UFI11/55

\nocite{Anthes_1982book, Emanuel_2003AnnRev}
\nocite{Nash_etal_1996_Earth-Polar-Vortex}
\nocite{Simon-Miller_etal_2002_GRS, Choi_etal_2007, Simon_etal_2014_GRS-Shrink}
\nocite{Piccioni_etal_2007VenSpole, Limaye_etal_2009VenusPolarVort, Patsaeva_etal_2015VenusWind}
\nocite{Orton_etal_2012NepPole, Sromovsky_etal_2001_PartIII_NepWind}
\nocite{Emanuel_2003AnnRev}
\nocite{Godfrey_1988}
\nocite{Sayanagi_etal_2015DPS}

\bibliography{bibliography}\label{refs}
\bibliographystyle{cambridgeauthordate-Leigh}
  


\end{document}